\documentstyle[aps,epsfig,float,amstex]{revtex}

\newcommand{\be}{\begin{equation}}
\newcommand{\ee}{\end{equation}}
\newcommand{\beq}{\begin{eqnarray}}
\newcommand{\eeq}{\end{eqnarray}}

\begin{document}
    
\def\gC{\mbox{\boldmath $C$}}
\def\gZ{\mbox{\boldmath $Z$}}
\def\gR{\mbox{\boldmath $R$}}
\def\gN{\mbox{\boldmath $N$}}
\def\ua{\uparrow}
\def\da{\downarrow}
\def\a{\alpha}
\def\b{\beta}
\def\g{\gamma}
\def\G{\Gamma}
\def\d{\delta}
\def\D{\Delta}
\def\e{\epsilon}
\def\ve{\varepsilon}
\def\z{\zeta}
\def\h{\eta}
\def\th{\theta}
\def\k{\kappa}
\def\l{\lambda}
\def\L{\Lambda}
\def\m{\mu}
\def\n{\nu}
\def\x{\xi}
\def\X{\Xi}
\def\p{\pi}
\def\P{\Pi}
\def\r{\rho}
\def\s{\sigma}
\def\S{\Sigma}
\def\t{\tau}
\def\f{\phi}
\def\vf{\varphi}
\def\F{\Phi}
\def\c{\chi}
\def\w{\omega}
\def\W{\Omega}
\def\Q{\Psi}
\def\q{\psi}
\def\de{\partial}
\def\inf{\infty}
\def\ra{\rightarrow}
\def\bra{\langle}
\def\ket{\rangle}
\title{ Analytic Correlation Functions of the two-dimensional Half Filled Hubbard
Model at Weak Coupling}

\author{Gianluca Stefanucci  and  Michele Cini}
\address{Istituto Nazionale di Fisica della Materia, Dipartimento di Fisica,\\
Universita' di Roma Tor Vergata, Via della Ricerca Scientifica, 1-00133\\
Roma, Italy}
\maketitle

\begin{abstract}

We derive explicit  spin and charge correlation functions of the $N \times N$ Hubbard model
from a recently obtained weak-coupling analytic ground state $|\Q^{[0]}_{AF}\ket$.
The spin correlation function shows an antiferromagnetic behaviour 
with different signs for the two sublattices and its Fourier tranform 
is peaked at ${\mathbf Q}=(\p,\p)$. The charge correlation function presents two 
valleys at 45 degrees from the axes. Both functions behave in a smooth 
way with increasing $N$;  
the results agree well with the available  numerical data.

\end{abstract}
  
\section{Introduction}
\label{intro}
{\small 

Let us consider the Hubbard model with hamiltonian 
\begin{equation}
H=H_{0}+\hat{W}=t\sum_{\s}\sum_{\bra {\mathbf r},{\mathbf r}'\ket}
c^{\dag}_{{\mathbf r}\s}c_{{\mathbf r}'\s}+
\sum_{{\mathbf r}}U\hat{n}_{{\mathbf r}\ua}\hat{n}_{{\mathbf r}\da},\;\;\;\;U>0,
\label{hamil}
\end{equation}
on a bipartite square lattice $\L={\cal A}\cup {\cal B}$ of $N\times N$ 
sites with periodic boundary 
conditions and even $N$. Here $\s=\ua,\da$ is the 
spin and ${\mathbf r},\;{\mathbf r}'$ the spatial  degrees of freedom of the creation and 
annihilation operators $c^{\dag}$ and $c$ respectively. The sum on 
$\bra {\mathbf r},{\mathbf r}'\ket$ is over the pairs of nearest neighbor sites and 
$\hat{n}_{{\mathbf r}\s}$  is the number operator on site 
${\mathbf r}$ of spin $\s$. The point symmetry is 
$C_{4v}$, the Group of a square\footnote{{\footnotesize
 $C_{4v}$ is the symmetry Group of a square. 
It is a finite Group of 
order 8 and it contains 4 one dimensional irreps, 
$A_{1},\;A_{2},\;B_{1},\;B_{2}$, and 1 two-dimensional one called $E$. 
The table of characters is
\begin{center}
    \vspace*{0.5 cm}
\begin{tabular}{|c|c|c|c|c|c|}
\hline 
$C_{4v}$ & $\mathbf{1}$ & $C_{2}$ & $C^{(+)}_{4},\,C^{(-)}_{4}$ & 
$\s_{x},\,\s_{y}$ & 
$\s'_{x},\,\s'_{y}$ \\
\hline 
$A_{1}$ & 1 & 1 & 1 & 1 & 1 \\
\hline 
$A_{2}$ & 1 & 1 & 1 & -1 & -1 \\
\hline 
$B_{1}$ & 1 & 1 & -1 & 1 & -1 \\
\hline 
$B_{2}$ & 1 & 1 & -1 & -1 & 1 \\
\hline 
$E$ & 2 & -2 & 0 & 0 & 0 \\
\hline 
\end{tabular}
\vspace*{0.5 cm}
\end{center}
}}; besides, $H$ is 
invariant under the  commutative Group of Translations ${\mathbf 
T}$ and hence under the Space Group ${\mathbf G}={\mathbf 
T} \otimes C_{4v} $; $\otimes$ means the semidirect product. 
The presence of spin and pseudospin symmetries\cite{lieb} leads to 
$SO_{4}$ Group\cite{ya}\cite{yaza}; below, we shall work in 
the  subspace of vanishing spin and 
pseudospin. 
We   represent sites by ${\mathbf r}=(x,y)$ and wave vectors by
${\mathbf k}=(k_{x},k_{y})=\frac{2\p}{N}(x,y)$,
with  $x,y=0,\ldots,N-1$. In terms of the 
Fourier expanded fermion operators 
$c_{{\mathbf k}\s}=\frac{1}{N}\sum_{{\mathbf r}}
e^{i{\mathbf k}\cdot {\mathbf r}}c_{{\mathbf r}\s}$, we have 
$H_{0}=\sum_{{\mathbf k}}\e({\mathbf k})c^{\dag}_{{\mathbf k}\s}
c_{{\mathbf k}\s}$ with $\e({\mathbf k})=2t
[\cos k_{x}+\cos k_{y}]$. Then the one-body plane wave state 
$c^{\dag}_{{\mathbf k}\s}|0\ket\equiv|{\mathbf k}\s\ket$ 
is an  eigenstate of $H_{0}$.

The $N \times N$ Hubbard model at half filling is not elementary, 
even  in the innocent-looking case of finite $N$ and 
small repulsion $U$. Indeed, weak-coupling expansions have long been 
known to be highly informative \cite{friedman},\cite{galan}. 
The reason is that the trivial 
$U=0$ case is $\left(\begin{array}{c} 2N-2 \\ N-1 
\end{array}\right)^{2}$ 
times degenerate, and so even for relatively 
small lattices one has to solve a big secular problem to see how the 
interaction resolves the degeneracy in first-order. To deal with this 
problem, recently\cite{jop2001}\cite{ssc2001} we have 
proposed a {\em local }  formalism, based on
diagonalizing the occupation number 
operators in the degenerate 
eigenstates of the kinetic energy $H_{0}$. We came out with an 
analytic singlet wave function  $|\Q^{[0]}_{AF}\ket$ which solves the secular 
problem and belongs to the ground eigenvalue of $H$. 
Under a lattice step translation it just flips spins 
({\em antiferromagnetic property}). 
Further we proved that it 
has vanishing momentum and its point symmetry is the same as the 
ground state symmetry established by Moreo and Dagotto\cite{md}.

We believe that  $|\Q^{[0]}_{AF}\ket$ clearly deserves further study: 
although it is an eigenstate 
only at the first
order in $U$, it must represent a good deal of the properties of the 
full ground state. 
Here we wish to show that the same {\em local } formalism that allows 
one to build $|\Q^{[0]}_{AF}\ket$,  
is also suitable to bring out 
some physics. It is clear that the
importance of analytic results is actually enhanced in the  age 
of  computers, since they can benchmark the numerical approximations.

The correlation functions are a popular tool to understand and 
visualize the 
structure and the physical properties of a given many-body state.
The definitions are:  
\begin{equation}
G_{{\mathrm charge}}({\mathbf r})\equiv\bra\Q^{[0]}_{AF}|\hat{n}_{{\mathbf r}}
\hat{n}_{0}|\Q^{[0]}_{AF}\ket,
\label{deffdccc}
\end{equation}
for  the charge  correlation function and
\begin{equation}
G_{{\mathrm spin}}({\mathbf r})\equiv\bra\Q^{[0]}_{AF}|\hat{{\mathbf S}}_{{\mathbf r}}
\cdot \hat{{\mathbf S}}_{0}|\Q^{[0]}_{AF}\ket.
\label{deffdcss}
\end{equation}
for the spin one. Here $\hat{n}_{{\mathbf r}}$ is the number operator 
and $\hat{{\mathbf S}}_{{\mathbf r}}$ is 
the spin vector operator at site ${\mathbf r}$; 
the pedices 0 denote the site at the origin.
  
Studies\cite{ki} of correlation functions in the three-band
Hubbard model have been aimed to the characterization of possible 
pairing mechanisms short after the discovery of high-$T_{c}$ 
superconductivity\cite{bm}.
Much of the early work dealt with the one-band model at strong coupling. 
Let us mention the exact diagonalization study by Kaxiras and Manousakis\cite{km} 
on the $\sqrt{10} \times \sqrt{10}$ lattice, showing the 
antiferromagnetic order at half filling;  the diagrammatic approach 
by  Gebhard and  Vollhardt\cite{gv} used the Gutzwiller ansatz, mainly 
for the 1$d$ chain.
Correlation functions on larger lattices of the one-band \cite{qmc}, 
and, more recently, of the three-band Hubbard 
Model\cite{qmc2}\cite{qmc3} have been obtained by Quantum-Monte-Carlo 
methods. They have also been used to benchmark the self-consistent 
theory by Vilk et al.\cite{vct}, which is basically a generalized 
Random-Phase Approximation.

After summarizing the local formalism and the ground state solution in 
Section \ref{ground}, we go over in Section \ref{parth} to a new picture 
by a  particle-hole canonical transformation which is convenient to 
calculate the correlation functions. We derive the spin correlation 
function in Section \ref{spin} and the charge one in Section \ref{charge}. 
The results are discussed and compared with available data in 
Section \ref{comp}.

\section{The ground state at weak coupling}
\label{ground}

In order to estabilish some notations we need to review  the ground 
state formalism\cite{jop2001},  \cite{ssc2001}. Let ${\cal S}_{hf}$ 
denote the set (or shell) of the ${\mathbf k}$ wave vectors 
such that $\e({\mathbf k})=0$. 
At half filling ($N^{2}$ particles) for $U=0$ the ${\cal S}_{hf}$  shell
 is half occupied, while all $|{\mathbf k}\ket$ orbitals such that 
$\e({\mathbf k})<0$ are filled. The ${\mathbf k}$ vectors of 
${\cal S}_{hf}$ lie on the square having 
vertices $(\pm\pi,0)$ and $(0,\pm\pi)$;
one  readily realizes that the dimension of the 
set ${\cal S}_{hf}$, is $|{\cal S}_{hf}|=2N-2$. Since $N$ is even 
and $H$ commutes with the total spin  operators,
\begin{equation}
\hat{S}^{z}=\frac{1}{2}\sum_{{\mathbf r}}(\hat{n}_{{\mathbf r}\ua}-
\hat{n}_{{\mathbf r}\da}),\;\;\;\;
\hat{S}^{+}=\sum_{{\mathbf r}}c^{\dag}_{{\mathbf r}\ua}c_{{\mathbf r}\da},\;\;\;\;
\hat{S}^{-}=(\hat{S}^{+})^{\dag},
\label{su2gen}
\end{equation} 
at half filling every ground state of $H_{0}$ is represented in 
the $\hat{S}^{z}=0$ subspace. Thus, $H_{0}$ has  
$\left(\begin{array}{c} 2N-2 \\ N-1 
\end{array}\right)^{2}$   degenerate unperturbed ground state 
configurations with $\hat{S}^{z}=0$. 

It can be shown\cite{jop2001} that the structure of the first-order wave functions 
is gained by  diagonalizing $\hat{W}$ in the 
{\it  truncated  Hilbert space ${\cal H}$} spanned 
by the {\it states of $N-1$ holes of each spin in ${\cal S}_{hf}$}. 
In other terms, one solves a $(2N-2)$-particle problem in the truncated  
Hilbert space ${\cal H}$ and then, understanding the particles in the 
filled shells, obtains the first-order eigenfunctions of $H$ in the 
full $N^{2}$-particle problem. We emphasize that the matrix of $H_{0}$ 
in ${\cal H}$ is null,  since 
by construction ${\cal H}$ is contained in the kernel of $H_{0}$.

The large set  ${\cal S}_{hf}$ breaks into small pieces if we take full advantage
of the $\mathbf{G}$ symmetry. Any plane-wave state ${\mathbf k}$ belongs to a 
one-dimensional irrep of $\mathbf{T}$; moreover, it also belongs to 
a {\em star} of ${\mathbf k}$
vectors connected by operations of $C_{4v}$, and one member of the star 
has $k_{x}\geq k_{y}\geq 0$. We recall that any ${\mathbf k}\in {\cal S}_{hf}$  
lies on a square with vertices on the axes at the Brillouin zone 
boundaries.  Choosing an arbitrary ${\mathbf k}\in {\cal S}_{hf}$ with
$k_{x}\geq k_{y}\geq 0$, hence $k_{x}+k_{y}=\pi$, the set of vectors 
$R_{i}{\mathbf k} \in {\cal S}_{hf}$,  
where $R_{i}\in C_{4v}$, is a basis for an irrep of $\mathbf{G}$.
The high symmetry vectors ${\mathbf k}_{A}=(\pi,0)$ and ${\mathbf k}_{B}=(0,\pi)$ 
 are the basis of the only two-dimensional irrep of $\mathbf{G}$, which exists for any $N$.
If $N/2$ is even,  one also finds the high symmetry  wavevectors 
${\mathbf k}=(\pm\pi/2,\pm\pi/2)$ which mix among themselves under 
$C_{4v}$ operations and yield
a four-dimensional irrep.  In general, when ${\mathbf k}$ is not in a special 
symmetry direction,  the vectors $R_{i}{\mathbf k}$ are all 
different, so  all the other irreps of $\mathbf{G}$ have dimension 8, 
the number of operations of the point Group $C_{4v}$. 

Below, we shall need the number of these irreps.  Since 8 times the number of 
eight-dimensional irreps + 4 times that of four-dimensional ones + 2 
for the only two-dimensional irrep 
must yield $|{\cal S}_{hf}|=2N-2$, one finds that ${\cal S}_{hf}$ contains 
$N_{e}=\frac{1}{2}(\frac{N}{2}-2)$ irreps of dimension 8 if $N/2$ is even and $N_{o}
=\frac{1}{2}(\frac{N}{2}-1)$ irreps of dimension 8 if $N/2$ is odd. 

In this way, ${\cal S}_{hf}$ is seen to be the union of disjoint 
bases of irreps of the Space Group. This break-up of  ${\cal 
S}_{hf}$ enables us to define a {\em real} symmetry adapted  one-body {\em local} 
basis which allows to carry on the analysis for any $N$.  

The one-body {\em local} basis is obtained by projecting  onto the 
irreps of $C_{4v}$ the $|{\mathbf k}\ket$ states 
of ${\cal S}_{hf}$ that belong to a given irrep of $\mathbf{G}$. As already noted,  
${\mathbf k}_{A}=(\p,0)$ and ${\mathbf k}_{B}=(0,\p)$ belong to ${\cal S}_{hf}$ 
and are the basis of a two-dimensional irrep  of $\mathbf{G}$. Let 
\begin{equation}
|\q^{''}_{A_{1}}\ket=\frac{1}{\sqrt{2}}(|{\mathbf k}_{A}\ket+|{\mathbf k}_{B}\ket),\;\;\;\;
|\q^{''}_{B_{1}}\ket=\frac{1}{\sqrt{2}}(|{\mathbf k}_{A}\ket-|{\mathbf k}_{B}\ket)
\end{equation}
be the first two real states of the local basis. 
As the notation implies, both are simultaneously eigenvectors of the 
Dirac characters of $C_{4v }$ and carry symmetry labels; actually the symmetries 
are $A_{1}$ and $B_{1}$ because the two-dimensional irrep  of $\mathbf{G}$
breaks into $A_{1}\oplus B_{1}$ in $C_{4v }$. In $\mathbf{G}$ these 
two functions merge into one irrep because the ${\mathbf k}$ states pick up 
phase factors from the translations.

For even $N/2$, ${\cal S}_{hf}$ also comprises the basis wave vectors 
${\mathbf k}_{1}=(\p/2,\p/2),\;{\mathbf k}_{2}=(-\p/2,\p/2),\;{\mathbf k}_{3}=(\p/2,-\p/2),\; 
{\mathbf k}_{4}=(-\p/2,-\p/2)$ of the 4-dimensional irrep of $\mathbf{G}$. This 
irrep  breaks into $A_{1}\oplus B_{2}\oplus 
E$ in $C_{4v }$. Letting $I=1,2,3,4$ for the irreps $A_{1},\;B_{2},\;E_{x},\;E_{y}$ 
respectively, we can  write down  four more real local states 
\begin{equation}
|\q^{'}_{I}\ket=\sum_{i=1}^{4}O'_{Ii}|{\mathbf k}_{i}\ket,
\end{equation}
where $O'$ is the  $4\times 4$ unitary matrix which performs the 
projections, namely,
\begin{equation}
O'=\frac{1}{2}\left[\begin{array}{rrrr}
1 & 1 & 1 & 1 \\
1 & -1 & -1 & 1 \\
i & -i & i & -i \\
i & i & -i & -i \end{array}\right].
\end{equation}
For $N>4$, ${\cal S}_{hf}$ also contains ${\mathbf k}$ vectors that are away 
from  special symmetry directions.  These form  eight-dimensional irreps of ${\mathbf G}$  
since  $R_{i}{\mathbf k}$ are all different for all $R_{i} \in C_{4v}$.  
In other terms, any eight-dimensional irrep of ${\mathbf G}$ is the regular representation 
of $C_{4v}$.  Thus, by the Burnside theorem, it breaks  into 
$A_{1}\oplus A_{2}\oplus B_{1}\oplus B_{2}\oplus E\oplus E$, with the 
two-dimensional irrep occurring twice; these are the symmetry labels 
of the local orbitals we are looking for. Let ${\mathbf k}^{[m]}=
(k^{[m]}_{x},k^{[m]}_{y})$ with $k^{[m]}_{x}\geq k^{[m]}_{y}\geq 0$ be a wave vector of 
the $m$-th eight dimensional irrep of ${\mathbf G}$ and let 
$R_{i},\;i=1,\ldots,8$ denote respectively the identity $\mathbf{1}$, the 
counterclockwise and clockwise 90 degrees rotation
 $C_{4}^{(+)},\;C_{4}^{(-)}$, 
the  180 degrees rotation $C_{2}$, the reflection with respect 
to the $y=0$ and $x=0$ axis $\s_{x},\;\s_{y}$ 
and the reflection with respect to the 
$x=y$ and $x=-y$ diagonals $\s'_{x},\;\s'_{y}$.  
We write down real local basis states as 
\begin{equation}
    |\q^{[m]}_{I}\ket=\sum_{i=1}^{8}O_{Ii}|R_{i}{\mathbf k}^{[m]}\ket,
\end{equation}
where $O$ is the 8$\times$8 unitary matrix 
\begin{equation}
O=\frac{1}{\sqrt{8}}
\left[\begin{array}{rrrrrrrr}
                              1 & 1 & 1 & 1 & 1 & 1 & 1 & 1 \\
			      1 & -1 & -1 & 1 & -1 & -1 & 1 & 1 \\
			      i & i & -i & -i & i & -i & -i & i \\
			      i & -i & i & -i & -i & i & -i & i \\
                              1 & 1 & 1 & 1 & -1 & -1 & -1 & -1 \\
			      1 & -1 & -1 & 1 & 1 & 1 & -1 & -1 \\
			      i & -i & i & -i & i & -i & i & -i \\
			      i & i & -i & -i & -i & i & i & -i 
\end{array}\right].
\label{ort}
\end{equation}
Here, denoting by $E^{\prime}$ the second occourrence of the irrep $E$,
$I=1,\ldots,8$  is the
$A_{1}$, $B_{2}$, $E_{x}$, $E_{y}$, $A_{2}$, 
$B_{1}$, $E^{\prime}_{x}$, $E^{\prime}_{y}$ 
irrep respectively.

Now let us consider the following determinantal state
\begin{equation}
|\F_{AF}\ket_{\s}\equiv|(\prod_{m=1}^{N_{e}}
\q^{[m]}_{A_{1}}\q^{[m]}_{B_{2}}\q^{[m]}_{E_{x}}\q^{[m]}_{E_{y}})
\q'_{A_{1}}\q'_{B_{2}}\q''_{A_{1}}
\ket_{\s}\otimes 
|(\prod_{m=1}^{N_{e}}\q^{[m]}_{A_{2}}\q^{[m]}_{B_{1}}\q^{[m]}_{E'_{x}}
\q^{[m]}_{E'_{y}})\q'_{E_{x}}\q'_{E_{y}}\q''_{B_{1}}\ket_{-\s},
\label{detaf}
\end{equation}
for even $N/2$ and 
\begin{equation}
|\F_{AF}\ket_{\s}\equiv|(\prod_{m=1}^{N_{o}}
\q^{[m]}_{A_{1}}\q^{[m]}_{B_{2}}\q^{[m]}_{E_{x}}\q^{[m]}_{E_{y}})
\q''_{A_{1}}
\ket_{\s}\otimes 
|(\prod_{m=1}^{N_{o}}\q^{[m]}_{A_{2}}\q^{[m]}_{B_{1}}\q^{[m]}_{E'_{x}}
\q^{[m]}_{E'_{y}})\q''_{B_{1}}\ket_{-\s}.
\label{detafodd}
\end{equation}
for odd $N/2$, with $\s=\ua,\da$. In Ref.\cite{jop2001}\cite{ssc2001}
we have shown that
\begin{itemize}

\item
$|\F_{AF}\ket_{\s}$ is an eigenstate of $\hat{W}$ with vanishing 
eigenvalue ({\it $W=0$ state}).

\item
Under a lattice step translation $|\F_{AF}\ket_{\s}\ra - 
|\F_{AF}\ket_{-\s}$. Therefore, it manifestly shows an 
antiferromagnetic order ({\em antiferromagnetic property}). 

\item
Introducing the projection operator 
$P_{S}$  on the spin $S$ subspace, one finds that  
$P_{S}|\F_{AF}\ket_{\s}\equiv|\F^{[S]}_{AF}\ket\neq 0,\;\forall S=0,\ldots,N-1$. 
Then, $\bra \F_{AF}|\hat{W}|\F_{AF}\ket=\sum_{S=0}^{N-1}\, 
\bra\F^{[S]}_{AF}|\hat{W}|\F^{[S]}_{AF}
\ket=0$, and this implies that there is at least one  $W=0$ state 
of $\hat{W}$ in  ${\cal H}$ for each $S$.
By the Lieb Theorem\cite{lieb},  only the singlet component $|\F^{[0]}_{AF}\ket$ belongs to 
the ground state multiplet of $H$ at weak coupling 
(filled shells are understood, of course).

\item
$|\F^{[0]}_{AF}\ket$ has vanishing total momentum, is even under 
reflections, while the point symmetry is $s$ or $d$ for even or odd 
$N/2$, respectively. These are the correct quantum numbers of the 
interacting ground state at half filling\cite{md} .

\item
The ground state interaction energy per site is 
\begin{equation}
E_{U}\equiv\frac{\bra\F^{[0]}_{AF}|\hat{W}|\F^{[0]}_{AF}\ket}{N^{2}}=
\frac{U}{4}-U\frac{(N-1)^{2}}{N^{4}}.
\label{gse}
\end{equation}
Thus  the linear term of the expansion 
of the energy per site in powers of $U$ increases monotonically with 
$N$ towards  the infinite square lattice value $U/4$.

\end{itemize}

Finally we emphasize that in $|\F^{[0]}_{AF}\ket$ only $2N-2$ particles are 
antiferromagnetically correlated while in the strong coupling limit 
all the $N^{2}$ particles show antiferromagnetic correlations.

$|\F^{[0]}_{AF}\ket$ is an exact ground state of $H$ for  $U\ra 0$. 
Does this mean that it is  the $U\ra 0$ limit 
of the unique interacting ground state? We know that this is the case 
for the $4\times 4$ and $6\times 6$ square  
lattices, where we have numerical evidence that  $|\F^{[0]}_{AF}\ket$  is the only singlet 
eigenstate in ${\cal H}$ with vanishing eigenvalue. 
The correlation functions that we find below behave quite 
reasonably also for $N>6$ and this strongly suggests that  
$|\F^{[0]}_{AF}\ket$ continues to be a good approximation to the true 
ground state at weak coupling.
A proof of the uniqueness  of the vanishing eigenvalue of $\hat{W}$ in the singlet 
subspace of ${\cal H}$ would be sufficient (although not necessary) to 
prove that.
In the next Section we find further evidence to support 
this proposal: we write $|\F^{[0]}_{AF}\ket$ in the particle-hole transformed 
picture and show that the corresponding Lieb matrix is positive 
semidefinite, as it should be for a genuine ground state\cite{lieb}.

\section{The ground state in the particle-hole transformed picture}
\label{parth}

The unitary particle-hole transformation on a square $N\times N$ lattice and 
even $N$ reads
\begin{equation}
\left\{\begin{array}{l}
c_{{\mathbf r}\da}=d_{{\mathbf r}\da} \\ c_{{\mathbf r}\ua}=
(-)^{x+y}d^{\dag}_{{\mathbf r}\ua}
\end{array}\right. 
,\;\;\;\;\;\;\;\;\;\;{\mathbf r}=(x,y).
\label{upht}
\end{equation}
This transformation maps the repulsive Hubbard model described in Eq.(\ref{hamil}) 
onto the attractive one
\begin{equation}
H=t\sum_{\s}\sum_{\bra {\mathbf r},{\mathbf r}'\ket}
d^{\dag}_{{\mathbf r}\s}d_{{\mathbf r}'\s}-
\sum_{{\mathbf r}}U\hat{n}^{(d)}_{{\mathbf r}\ua}\hat{n}^{(d)}_{{\mathbf r}\da}+
U\hat{N}^{(d)}_{\da},\;\;\;\;U>0,
\label{hamilpht}
\end{equation}
with $\hat{n}^{(d)}_{{\mathbf r}\s}=d^{\dag}_{{\mathbf r}\s}d_{{\mathbf r}\s}$ and 
$\hat{N}^{(d)}_{\s}=\sum_{{\mathbf r}}\hat{n}^{(d)}_{{\mathbf r}\s}$. Letting 
$\{|\Q_{\G}\ket\}$ be an orthonormal real basis of $N^{2}/2$-particle 
states (that is, each $|\Q_{\G}\ket$ must be an homogeneous 
polynomial of degree $N^{2}/2$ in the $d^{\dag}_{{\mathbf r}}$ with real 
coefficients acting on the vacuum), we remind that the ground state at 
half filling 
\begin{equation}
|\Q^{[0]}\ket=\sum_{\G_{1}\G_{2}}L_{\G_{1}\G_{2}}|\Q_{\G_{1}\ua}\ket
\otimes|\Q_{\G_{2}\da}\ket
\end{equation}
is such that the Lieb matrix $L_{\G_{1}\G_{2}}$ is positive (or 
negative) semidefinite. 

In this Section we will perform the unitary particle-hole 
transformation in Eq.(\ref{upht}) on the ground state of 
Eqs.(\ref{detaf})(\ref{detafodd}). We will show that the corresponding 
Lieb matrix is indeed positive semidefinite; besides, it is already diagonal in the local 
basis. As a consequence, the ground state in the untransformed ($c$)
picture is a pseudospin  (as well as spin) singlet.

Let $c^{\dag}_{i}$, $i=1,\ldots,2N-2$, be the operators which create 
a particle in the $i$-th local state contained in $|\F_{AF}\ket_{\s}$, 
Eqs.(\ref{detaf})(\ref{detafodd}). We write $|\F_{AF}\ket_{\s}$ as 
\begin{equation}
|\F_{AF}\ket_{\s}=c^{\dag}_{1,\s}\ldots 
c^{\dag}_{N-1,\s}c^{\dag}_{N,-\s}\ldots 
c^{\dag}_{2N-2,-\s}|0\ket.
\end{equation}
It is clear that the first [last] $N-1$ creation operators refer to 
the states of spin $\s$ [$-\s$] in Eqs.(\ref{detaf})(\ref{detafodd}).
The singlet projection gives
\begin{eqnarray}
|\F^{[0]}_{AF}\ket=P_{S=0}|\F_{AF}\ket_{\ua}=\frac{1}{\sqrt{{\cal 
N}}}
\{\sum_{k=N/2}^{1}(-)^{k}g_{k}\sum_{i_{\frac{N}{2}-k}>..>i_{1}=1}^{N-1}
\hat{S}^{-}_{i_{1}}..\hat{S}^{-}_{i_{\frac{N}{2}-k}}
\sum_{j_{\frac{N}{2}-k}>..>j_{1}=N}^{2N-2}\hat{S}^{+}_{j_{1}}..
\hat{S}^{+}_{j_{\frac{N}{2}-k}}\}\times\nonumber \\ \times 
c^{\dag}_{1,\ua}\ldots 
c^{\dag}_{N-1,\ua}c^{\dag}_{N,\da}\ldots 
c^{\dag}_{2N-2,\da}|0\ket+\nonumber \\ +\frac{1}{\sqrt{{\cal 
N}}}
\{\sum_{k=N/2}^{1}(-)^{k+1}g_{k}\sum_{i_{\frac{N}{2}-k}>..>i_{1}=1}^{N-1}
\hat{S}^{+}_{i_{1}}..\hat{S}^{+}_{i_{\frac{N}{2}-k}}
\sum_{j_{\frac{N}{2}-k}>..>j_{1}=N}^{2N-2}\hat{S}^{-}_{j_{1}}..
\hat{S}^{-}_{j_{\frac{N}{2}-k}}\}\times\nonumber \\ \times 
c^{\dag}_{1,\da}\ldots 
c^{\dag}_{N-1,\da}c^{\dag}_{N,\ua}\ldots 
c^{\dag}_{2N-2,\ua}|0\ket
\end{eqnarray}
where $\hat{S}^{+}_{i}=c^{\dag}_{i,\ua}c_{i,\da}$, 
$\hat{S}^{-}_{i}=(\hat{S}^{+}_{i})^{\dag}$, 
${\cal N}$ is the normalization constant 
\begin{equation}
{\cal N}=2\cdot\sum_{k=1}^{N/2}g_{k}^{2}\left(\begin{array}{c}
N-1 \\ N/2-k \end{array}\right)^{2}
\end{equation}
and the $g_{k}$'s are given by 
\begin{equation}
g_{k}=\frac{
\left(\begin{array}{c}
N/2+k-1 \\ N/2-1 \end{array}\right)
}{
\left(\begin{array}{c}
N/2 \\ N/2-k \end{array}\right)
}.
\end{equation}
Let
\begin{equation}
c_{{\mathbf k}}=\frac{1}{N}\sum_{{\mathbf r}}
e^{i{\mathbf k}\cdot {\mathbf r}}c_{{\mathbf r}},\;\;\;\;\;
d_{{\mathbf k}}=\frac{1}{N}\sum_{{\mathbf r}}
e^{i{\mathbf k}\cdot {\mathbf r}}d_{{\mathbf r}},
\end{equation}
be the Fourier transformed operators of the site annihilation 
operators $c_{{\mathbf r}}$ and $d_{{\mathbf r}}$ respectively. From 
Eq.(\ref{upht}) we get
\begin{equation}
c_{{\mathbf k}\da}=d_{{\mathbf k}\da},\;\;\;\;\;
c_{{\mathbf k}\ua}=d^{\dag}_{{\mathbf Q}-{\mathbf k}\ua},\;\;\;\;\;\;\;\;\;
{\mathbf Q}=(\p,\p).
\end{equation}
The ground state with the Fermi sea explicitly written is given by 
$|\Q^{[0]}_{AF}\ket=|\F^{[0]}_{AF}\ket\otimes|\S\ket$ 
where $|\S\ket$ is the contribution 
from the filled shells:
\begin{equation}
|\S\ket=|\S_{\ua}\ket\otimes|\S_{\da}\ket,\;\;\;\;
|\S_{\s}\ket=\prod_{\e({\mathbf k})<0}c^{\dag}_{{\mathbf k}\s}|0\ket.
\end{equation}
Modulo an overall phase factor, the particle-hole transformation  
yields 
\begin{equation}
|\S_{\da}\ket=\prod_{\e({\mathbf k})<0}c^{\dag}_{{\mathbf k}\da}|0\ket=
\prod_{\e({\mathbf k})<0}d^{\dag}_{{\mathbf k}\da}|0\ket\equiv|\tilde{\S}_{\da}\ket
\end{equation}
\begin{equation}
|\S_{\ua}\ket=\prod_{\e({\mathbf k})<0}c^{\dag}_{{\mathbf k}\ua}|0\ket=
\prod_{\e({\mathbf k})<0}d_{{\mathbf Q}-{\mathbf k}\ua}
\prod_{{\mathbf k}}d^{\dag}_{{\mathbf k}\ua}|0\ket.
\end{equation}
Let $d_{i}$ be the operator obtained substituing 
$c_{{\mathbf k}}$ with $d_{{\mathbf k}}$ in 
the definition of $c_{i}$. We note that $\e({\mathbf k})<0$ corresponds 
to $\e({\mathbf Q}-{\mathbf k})>0$. 
Then, the spin up filled shells state 
$|\S_{\ua}\ket$ can be written as
\begin{equation}
|\S_{\ua}\ket=\prod_{\e({\mathbf k})\leq 0}d^{\dag}_{{\mathbf k}\ua}|0\ket=
d^{\dag}_{1,\ua}..d^{\dag}_{N-1,\ua}d^{\dag}_{N,\ua}..d^{\dag}_{2N-2,\ua}
\prod_{\e({\mathbf k})< 0}d^{\dag}_{{\mathbf k}\ua}|0\ket\equiv 
d^{\dag}_{1,\ua}..d^{\dag}_{N-1,\ua}d^{\dag}_{N,\ua}..d^{\dag}_{2N-2,\ua}
|\tilde{\S}_{\ua}\ket
\end{equation}
and hence
\begin{equation}
|\S_{\ua}\ket\otimes|\S_{\da}\ket=
d^{\dag}_{1,\ua}..d^{\dag}_{N-1,\ua}d^{\dag}_{N,\ua}..d^{\dag}_{2N-2,\ua}
|\tilde{\S}_{\ua}\ket\otimes|\tilde{\S}_{\da}\ket.
\end{equation}
The next step is to express $c_{i\s}$ in terms of $d_{i\s}$. 
By direct inspection one readly realizes that 
\begin{equation}
    c_{i\da}=d_{i\da}\;\;\;\forall i,\;\;\;\;\;\;\;\;\;\;\;\;\;\;\;\;\;\;
    c_{i\ua}=\left\{\begin{array}{rl}
    d^{\dag}_{i\ua} & \;\;\;\;i=1,\ldots,N-1 \\
    -d^{\dag}_{i\ua} & \;\;\;\;i=N,\ldots,2N-2
    \end{array}\right. \;.
\end{equation}
The above result implies that the raising operators $\hat{S}^{+}_{i}$ 
in the $d$ picture are given by 
\begin{equation}
\hat{S}^{+}_{i}=\left\{\begin{array}{ll}
    d_{i\ua}d_{i\da}\equiv D_{i} & \;\;\;\;i=1,\ldots,N-1 \\
    -d_{i\ua}d_{i\da}\equiv -D_{i} & \;\;\;\;i=N,\ldots,2N-2
    \end{array}\right.\; .
\end{equation}
These last three equations allow to rewrite the whole ground state 
$|\Q^{[0]}_{AF}\ket=|\F^{[0]}_{AF}\ket\otimes|\S\ket$ in the new picture:
\begin{eqnarray}
|\Q^{[0]}_{AF}\ket=\frac{1}{\sqrt{{\cal 
N}}}
\{\sum_{k=N/2}^{1}g_{k}\sum_{i_{\frac{N}{2}-k}>..>i_{1}=1}^{N-1}
D^{\dag}_{i_{1}}..D^{\dag}_{i_{\frac{N}{2}-k}}
\sum_{j_{\frac{N}{2}-k}>..>j_{1}=N}^{2N-2}D_{j_{1}}..
D_{j_{\frac{N}{2}-k}}\}\times\nonumber \\ \times 
d_{1,\ua}..d_{N-1,\ua}d^{\dag}_{N,\da}..d^{\dag}_{2N-2,\da}
d^{\dag}_{1,\ua}..d^{\dag}_{N-1,\ua}d^{\dag}_{N,\ua}..d^{\dag}_{2N-2,\ua}
|\tilde{\S}_{\ua}\ket\otimes|\tilde{\S}_{\da}\ket+
\nonumber \\ +\frac{1}{\sqrt{{\cal 
N}}}
\{\sum_{k=N/2}^{1}g_{k}\sum_{i_{\frac{N}{2}-k}>..>i_{1}=1}^{N-1}
D_{i_{1}}..D_{i_{\frac{N}{2}-k}}
\sum_{j_{\frac{N}{2}-k}>..>j_{1}=N}^{2N-2}D^{\dag}_{j_{1}}..
D^{\dag}_{j_{\frac{N}{2}-k}}\}\times\nonumber \\ \times 
d^{\dag}_{1,\da}..d^{\dag}_{N-1,\da}d_{N,\ua}..d_{2N-2,\ua}
d^{\dag}_{1,\ua}..d^{\dag}_{N-1,\ua}d^{\dag}_{N,\ua}..d^{\dag}_{2N-2,\ua}
|\tilde{\S}_{\ua}\ket\otimes|\tilde{\S}_{\da}\ket=
\nonumber \\ =
\frac{1}{\sqrt{{\cal N}}}
\{\sum_{k=N/2}^{1}g_{k}\sum_{i_{\frac{N}{2}-k}>..>i_{1}=1}^{N-1}
\sum_{j_{\frac{N}{2}-k}>..>j_{1}=N}^{2N-2}
D^{\dag}_{i_{1}}..D^{\dag}_{i_{\frac{N}{2}-k}}
D_{j_{1}}..D_{j_{\frac{N}{2}-k}}\}\times\nonumber \\ \times 
d^{\dag}_{N,\ua}..d^{\dag}_{2N-2,\ua}
d^{\dag}_{N,\da}..d^{\dag}_{2N-2,\da}
|\tilde{\S}_{\ua}\ket\otimes|\tilde{\S}_{\da}\ket+
\nonumber \\ +\frac{1}{\sqrt{{\cal N}}}
\{\sum_{k=N/2}^{1}g_{k}\sum_{i_{\frac{N}{2}-k}>..>i_{1}=1}^{N-1}
\sum_{j_{\frac{N}{2}-k}>..>j_{1}=N}^{2N-2}
D_{i_{1}}..D_{i_{\frac{N}{2}-k}}
D^{\dag}_{j_{1}}..D^{\dag}_{j_{\frac{N}{2}-k}}\}\times\nonumber \\ \times 
d^{\dag}_{1,\ua}..d^{\dag}_{N-1,\ua}
d^{\dag}_{1,\da}..d^{\dag}_{N-1,\da}
|\tilde{\S}_{\ua}\ket\otimes|\tilde{\S}_{\da}\ket.
\end{eqnarray}
Therefore, the singlet ground state has the following form
\begin{equation}
|\Q^{[0]}_{AF}\ket=
\sum_{\G}w_{\G}{\cal D}^{\dag}_{\G\ua}
|\tilde{\S}_{\ua}\ket\otimes {\cal D}^{\dag}_{\G\da}|\tilde{\S}_{\da}\ket,
\label{gsapht}
\end{equation}
where $\G=\{\g_{1},..,\g_{N-1}\}$ with 
$1\leq \g_{1}<..<\g_{N-1}\leq 2N-2$ and ${\cal D}^{\dag}_{\G}=
d^{\dag}_{\g_{1}}..d^{\dag}_{\g_{N-1}}$. $w_{\G}$ is the 
amplitude corresponding to the configuration $\G$. If  
in $\G$ there are $p$ indices between 1 and $N-1$ and $N-1-p$ indices 
between $N$ and $2N-2$, or {\em viceversa}, the amplitude 
$w_{\G}$ is given by
\begin{equation}
w_{\G}= \frac{1}{\sqrt{{\cal N}}}g_{N/2-p}.
\end{equation}

We conclude that the Lieb matrix is already diagonal in the local 
basis, and the nonvanishing diagonal elements are  $w_{\G}>0$. Thus it 
is positive semidefinite, which is consistent with its use as the 
ground state. We are now in position to calculate the correlation 
functions.

\section{The Spin Correlation Function}
\label{spin}      
In this Section we will explicitly write down an exact analytic 
formula for the spin correlation function of the half 
filled Hubbard model in the limit of vanishing interaction. In 
particular we evaluate
\begin{equation}
G_{\mathrm{spin}}({\mathbf r})=\bra\Q^{[0]}_{AF}|
{\mathbf S}_{{\mathbf r}}\cdot{\mathbf S}_{0}|\Q^{[0]}_{AF}\ket,
\;\;\;\;\;
{\mathbf S}_{0}\equiv {\mathbf S}_{{\mathbf r}=(0,0)}
\end{equation}
where ${\mathbf S}_{{\mathbf r}}=(\hat{S}^{x}_{{\mathbf r}},
\hat{S}^{y}_{{\mathbf r}},\hat{S}^{z}_{{\mathbf r}})$ is 
the spin vector operator at site ${\mathbf r}$ with components  
\begin{equation}
\hat{S}^{x}_{{\mathbf r}}=\frac{1}{2}(\hat{S}^{+}_{{\mathbf r}}+\hat{S}^{-}_{{\mathbf r}}), 
\;\;\;\;
\hat{S}^{y}_{{\mathbf r}}=\frac{1}{2i}(\hat{S}^{+}_{{\mathbf r}}-\hat{S}^{-}_{{\mathbf r}}),
\;\;\;\;
\hat{S}^{z}_{{\mathbf r}}=\frac{1}{2}(\hat{n}_{{\mathbf r}\ua}-\hat{n}_{{\mathbf r}\da})
\end{equation}
and 
\begin{equation}
\hat{S}^{+}_{{\mathbf r}}=c^{\dag}_{{\mathbf r}\ua}c_{{\mathbf r}\da},\;\;
\hat{S}^{-}_{{\mathbf r}}=(\hat{S}^{+}_{{\mathbf r}})^{\dag}=
c^{\dag}_{{\mathbf r}\da}c_{{\mathbf r}\ua}.
\end{equation}
Taking into account that $\Q^{[0]}_{AF}$ is a singlet, one has 
$\bra \hat{S}^{x}_{{\mathbf r}}\hat{S}^{x}_{0}\ket=
\bra \hat{S}^{y}_{{\mathbf r}}\hat{S}^{y}_{0}\ket=
\bra\hat{S}^{z}_{{\mathbf r}}\hat{S}^{z}_{0}\ket$ and hence
\begin{equation}
G_{\mathrm{spin}}({\mathbf r})=\frac{3}{4}[
\bra\hat{S}^{-}_{{\mathbf r}}\hat{S}^{+}_{0}+
\hat{S}^{+}_{{\mathbf r}}\hat{S}^{-}_{0}\ket]
\end{equation}
with $\hat{S}^{\pm}_{0}=\hat{S}^{\pm}_{{\mathbf r}=(0,0)}$ and 
$\bra\ldots\ket$  means the 
expectation value over the ground state $|\Q^{[0]}_{AF}\ket$.
Since $|\Q^{[0]}_{AF}\ket$ is a real linear combination of real basis 
vectors, $G^{-+}({\mathbf r})\equiv\bra\hat{S}^{-}_{{\mathbf r}}\hat{S}^{+}_{0}\ket 
\in \Re \;\;\forall {\mathbf r}$ and this implies 
\begin{equation}
G^{-+}({\mathbf r})=G^{-+}({\mathbf r})^{\ast}=
\bra\hat{S}^{-}_{0}\hat{S}^{+}_{{\mathbf r}}\ket=
\bra\hat{S}^{+}_{{\mathbf r}}\hat{S}^{-}_{0}\ket-2\d_{{\mathbf r},0}
\bra\hat{S}^{z}_{{\mathbf r}}\ket.
\end{equation}
Noting that $|\Q^{[0]}_{AF}\ket$ is a translation invariant state, 
$\bra\hat{S}^{z}_{0}\ket=\frac{1}{N^{2}}\sum_{{\mathbf r}}
\bra\hat{S}^{z}_{{\mathbf r}}\ket=
\frac{1}{N^{2}}\bra\hat{S}^{z}\ket=0$ and hence
\begin{equation}
G^{-+}({\mathbf r})=\bra\hat{S}^{+}_{{\mathbf r}}
\hat{S}^{-}_{0}\ket\equiv G^{+-}({\mathbf r}).
\end{equation}
This last equation allow us to express the spin correlation function 
$G_{\mathrm{spin}}({\mathbf r})$ in terms of $G^{-+}({\mathbf r})$ only 
\begin{equation}
G_{\mathrm{spin}}({\mathbf r})=\frac{3}{2}G^{-+}({\mathbf r}),
\label{gspindigmp}
\end{equation}
and the original problem is reduced to the calculation of $G^{-+}({\mathbf r})$. 
In the following we will show that $G^{-+}({\mathbf r})$ can be expressed in 
terms of three main contributions, two of which easily computable in 
the particle-hole transformed picture. Therefore, it is convenient to 
express $G^{-+}({\mathbf r})$ in terms of the $d$'s operators:
\begin{eqnarray}
G^{-+}({\mathbf r})=\bra c^{\dag}_{{\mathbf r}\da}
c_{{\mathbf r}\ua}c^{\dag}_{0\ua}c_{0\da}\ket=
(-)^{x+y}\bra d^{\dag}_{{\mathbf r}\da}d^{\dag}_{{\mathbf r}\ua}d_{0\ua}d_{0\da}\ket=
(-)^{x+y}\sum_{\G_{1}\G_{2}}w_{\G_{1}}w_{\G_{2}}
\bra\tilde{\S}_{\da}|{\cal D}_{\G_{1}\da}d^{\dag}_{{\mathbf r}\da}d_{0\da}
{\cal D}^{\dag}_{\G_{2}\da}|\tilde{\S}_{\da}\ket\times\nonumber \\
\times 
\bra\tilde{\S}_{\ua}|{\cal D}_{\G_{1}\ua}d^{\dag}_{{\mathbf r}\ua}d_{0\ua}
{\cal D}^{\dag}_{\G_{2}\ua}|\tilde{\S}_{\ua}\ket\equiv\nonumber \\ \equiv
(-)^{x+y}\sum_{\G_{1}\G_{2}}w_{\G_{1}}w_{\G_{2}}
G_{\G_{1}\G_{2}}({\mathbf r})^{2},
\label{sqtth}
\end{eqnarray}
where, dropping the spin index, 
\begin{equation}
G_{\G_{1}\G_{2}}({\mathbf r})=\bra\G_{1}|
d^{\dag}_{{\mathbf r}}d_{0}|\G_{2}\ket,\;\;\;\;\;\;
|\G\ket={\cal D}^{\dag}_{\G}|\tilde{\S}\ket.
\end{equation}
Here and in the following $c_{0\s}\equiv c_{{\mathbf r}=(0,0)\s}$ and 
$d_{0\s}\equiv d_{{\mathbf r}=(0,0)\s}$. 
Since the $w_{\G}$'s are non-negative, Eq.(\ref{sqtth}) shows that 
$G^{-+}({\mathbf r})$ is positive if ${\mathbf r}$ belongs to the 
sublattice ${\cal A}$ containing ${\mathbf r}=(0,0)$ and negative otherwise. This 
was pointed out in Ref. \cite{sqt}.  
All the information on the spin correlation function is  
enclosed into the site-dependent matrix elements 
$G_{\G_{1}\G_{2}}({\mathbf r})$. To evaluate them we write  the annihilation 
operator $d_{{\mathbf r}}$ as the sum of three pieces 
\begin{equation}
d_{{\mathbf r}}=\frac{1}{N}\sum_{{\mathbf k}}
e^{i{\mathbf k}\cdot {\mathbf r}}d_{{\mathbf k}}=\frac{\sqrt{|{\cal 
S}_{hf}|}}{N}d_{1}({\mathbf r})+
\frac{\sqrt{|{\cal S}|}}{N}d_{\x}({\mathbf r})+\frac{\sqrt{|{\cal 
S}|}}{N}d_{\bar{\x}}({\mathbf r})
\equiv \r_{hf}d_{1}({\mathbf r})+\r[d_{\x}({\mathbf r})+d_{\bar{\x}}({\mathbf r})],
\label{deco}
\end{equation}
with
\begin{equation}
d_{1}({\mathbf r})=\frac{1}{\sqrt{|{\cal S}_{hf}|}}\sum_{\e({\mathbf k})=0}e^{i{\mathbf 
k}\cdot {\mathbf r}}d_{{\mathbf k}},\;\;\;\;\;\;\;
d_{\x}({\mathbf r})=\frac{1}{\sqrt{|{\cal S}|}}\sum_{\e({\mathbf k})<0}e^{i{\mathbf 
k}\cdot {\mathbf r}}d_{{\mathbf k}},\;\;\;\;\;\;\;
d_{\bar{\x}}({\mathbf r})=\frac{1}{\sqrt{|{\cal S}|}}\sum_{\e({\mathbf 
k})>0}e^{i{\mathbf k}\cdot {\mathbf r}}d_{{\mathbf k}}
\label{dtra}
\end{equation}
and $|{\cal S}_{hf}|=2N-2$, $|{\cal S}|=\frac{1}{2}(N^{2}-|{\cal 
S}_{hf}|)$. We observe that $d_{1}(0)=d_{1}({\mathbf r}=(0,0))$ 
belongs to $A_{1}$ and that it can 
be written as a real linear combination of all the $A_{1}$-symmetric local 
annhilation operators of the local basis. By a unitary transformation on this 
$A_{1}$-subspace we may arrange that $d_{1}(0)$ is the new $d_{1}$. 
Thus, from now on, the one-body local basis $\{d^{\dag}_{i}|0\ket,\; 
i=1,\ldots,2N-2\}$, is 
such that the set of the $A_{1}$-symmetric local states contains 
$d^{\dag}_{1}(0)|0\ket$ and $d^{\dag}_{1}|0\ket=d^{\dag}_{1}(0)|0\ket$. 

Taking Eq.(\ref{deco}) into account one can 
express $G_{\G_{1}\G_{2}}({\mathbf r})$ as the sum of two terms:
\begin{equation}
G_{\G_{1}\G_{2}}({\mathbf r})=\r^{2}_{hf}\bra\G_{1}|d^{\dag}_{1}({\mathbf r})d_{1}|\G_{2}\ket+
\r^{2}\bra\G_{1}|d^{\dag}_{\x}({\mathbf r})d_{\x}|\G_{2}\ket\equiv
\r^{2}_{hf}G^{[hf]}_{\G_{1}\G_{2}}({\mathbf r})+
\r^{2}G^{[\x]}_{\G_{1}\G_{2}}({\mathbf r}), 
\label{twoterms}
\end{equation}
with $d_{1}\equiv d_{1}(0)$ and $d_{\x}\equiv d_{\x}(0)\equiv d_{\x}({\mathbf r}=(0,0))$. 
$G^{[\x]}_{\G_{1}\G_{2}}({\mathbf r})$ can be easily evaluated:
\begin{equation}
G^{[\x]}_{\G_{1}\G_{2}}({\mathbf r})=\bra\G_{1}|d^{\dag}_{\x}({\mathbf r})d_{\x}|\G_{2}\ket=
\d_{\G_{1}\G_{2}}\sum_{\e({\mathbf k}_{1})<0}\sum_{\e({\mathbf 
k}_{2})<0}e^{-i{\mathbf k}_{1}\cdot {\mathbf r}}
\frac{1}{|{\cal 
S}|}\bra\tilde{\S}|d^{\dag}_{{\mathbf k}_{1}}d_{{\mathbf k}_{2}}|\tilde{\S}\ket=
\d_{\G_{1}\G_{2}}\frac{{\cal T}({\mathbf r})}{|{\cal S}|}
\label{gxsi}
\end{equation}
where ${\cal T}({\mathbf r})=\sum_{\e({\mathbf k})<0}
e^{-i{\mathbf k}\cdot {\mathbf r}}$ is the trace of the  
translation matrix in the Hilbert space spanned by the negative 
energies one-body plane-wave states. Taking into account 
Eqs.(\ref{twoterms})(\ref{gxsi}), $G^{-+}({\mathbf r})$ in Eq.(\ref{sqtth}) can 
be rewritten as
\begin{equation}
G^{-+}({\mathbf r})=(-)^{x+y}\{\r^{4}_{hf}\sum_{\G_{1}\G_{2}}w_{\G_{1}}w_{\G_{2}}
G^{[hf]}_{\G_{1}\G_{2}}({\mathbf r})^{2}+
2\r^{2}\r_{hf}^{2}\frac{{\cal T}({\mathbf r})}{|{\cal S}|}
\sum_{\G}w_{\G}^{2}G^{[hf]}_{\G\G}({\mathbf r})+
\r^{4}\frac{{\cal T}({\mathbf r})^{2}}{|{\cal S}|^{2}}
\sum_{\G}w^{2}_{\G}\}.
\label{gpmfs}
\end{equation}
By definition 
$\sum_{\G}w^{2}_{\G}=\bra\Q^{[0]}_{AF}|\Q^{[0]}_{AF}\ket=1$. To 
evaluate the diagonal matrix elements $G^{[hf]}_{\G\G}({\mathbf r})$ we need to 
use the antiferromagnetic property. In the local basis, the one-body 
translation matrix has an antidiagonal block form if $x+y$ is 
odd and hence a diagonal block form otherwise. Therefore $d_{1}({\mathbf r})$ 
can be expanded as
\begin{equation}
d_{1}({\mathbf r})=\left\{\begin{array}{lll}
\sum_{i=1}^{N-1}t_{i}({\mathbf r})d_{i}, & \;t_{i}({\mathbf r})\in\Re & \;\;x+y\;{\mathrm even} \\
 & & \\
\sum_{i=N}^{2N-2}t_{i}({\mathbf r})d_{i}, & \;t_{i}({\mathbf r})\in\Re & \;\;x+y\;{\mathrm odd} 
\end{array}\right.
\label{dunotra}
\end{equation}
and $G^{[hf]}_{\G\G}({\mathbf r})$ becomes
\begin{equation}
G^{[hf]}_{\G\G}({\mathbf r})=\bra\G|d^{\dag}_{1}({\mathbf r})d_{1}|\G\ket=
\left\{\begin{array}{ll} t_{1}({\mathbf r})\d_{1\g_{1}} & x+y\;{\mathrm even} \\
0 &  x+y\;{\mathrm odd} 
\end{array}\right.
\label{ghfgg}
\end{equation}
where $\g_{1}$ is the first index of the configuration $\G$ (we remind 
that $\G=\{\g_{1},\ldots,\g_{N-1}\}$ with 
$1\leq\g_{1}<\ldots<\g_{N-1}\leq 2N-2$). Substituting this result into 
Eq.(\ref{gpmfs}) we see we need to evaluate 
$\sum_{\G}w_{\G}^{2}\d_{1\g_{1}}$. This can be done by observing that 
$|\F^{[0]}_{AF}\ket$ can be rewritten in the $c$ picture as 
\begin{equation}
|\F^{[0]}_{AF}\ket=|\F^{[0]}_{\ua}\ket-|\F^{[0]}_{\da}\ket
\label{decosm}
\end{equation}
with
\begin{equation}
|\F^{[0]}_{\ua}\ket=\frac{1}{\sqrt{{\cal N}}}\sum_{k=0}^{N-2}(-)^{k}f_{k}
\sum_{i_{k}>..>i_{1}=2}^{N-1}\sum_{j_{k}>..>j_{1}=N}^{2N-2}
\hat{S}^{-}_{i_{k}}..\hat{S}^{-}_{i_{1}}\hat{S}^{+}_{j_{k}}..\hat{S}^{+}_{j_{1}}
c^{\dag}_{1,\ua}\ldots c^{\dag}_{N-1,\ua}c^{\dag}_{N,\da}\ldots c^{\dag}_{2N-2,\da}
|0\ket
\label{fizeroalto}
\end{equation}
\begin{equation}
|\F^{[0]}_{\da}\ket=\frac{1}{\sqrt{{\cal N}}}\sum_{k=0}^{N-2}(-)^{k}f_{k}
\sum_{i_{k}>..>i_{1}=2}^{N-1}\sum_{j_{k}>..>j_{1}=N}^{2N-2}
\hat{S}^{+}_{i_{k}}..\hat{S}^{+}_{i_{1}}\hat{S}^{-}_{j_{k}}..\hat{S}^{-}_{j_{1}}
c^{\dag}_{1,\da}\ldots c^{\dag}_{N-1,\da}c^{\dag}_{N,\ua}\ldots c^{\dag}_{2N-2,\ua}
|0\ket
\label{fizerobasso}
\end{equation}
and
\begin{equation}
f_{k}=\left\{\begin{array}{ll}
g_{N/2-k} & \;\;\;k=0,\ldots,N/2-1 \\
g_{k+1-N/2} & \;\;\;k=N/2,\ldots,N-1
\end{array}\right. .
\end{equation}
All the configurations contained in 
$|\F^{[0]}_{\da}\ket$ are such  
that in the particle-hole transformed picture 
$\d_{1\g_{1}}=1$. On the other hand, all the configurations contained 
in $|\F^{[0]}_{\ua}\ket$ are such that 
in the particle-hole transformed picture 
$\d_{1\g_{1}}=0$. Therefore
\begin{equation}
\sum_{\G}w^{2}_{\G}\d_{1\g_{1}}=\bra\F^{[0]}_{\da}|\F^{[0]}_{\da}\ket=
\bra\F^{[0]}_{\ua}|\F^{[0]}_{\ua}\ket=\frac{1}{2}.
\label{norm2}
\end{equation}
The second term in Eq.(\ref{gpmfs}) is totally determined once we 
know $t_{1}({\mathbf r})$. By definition
\begin{equation}
t_{1}({\mathbf r})=\bra 0|d^{\dag}_{1}({\mathbf r})d_{1}|0\ket=
\frac{{\cal T}_{hf}({\mathbf r})}{|{\cal S}_{hf}|},
\label{t1dir}
\end{equation}
where ${\cal T}_{hf}({\mathbf r})=\sum_{\e({\mathbf k})=0}
e^{i{\mathbf k}\cdot {\mathbf r}}$ is the trace of the  
translation matrix in the Hilbert space spanned by the   
$\e({\mathbf k})=0$ one-body plane-wave states. In the last equalility of 
Eq.(\ref{t1dir}) we have used Eq.(\ref{dtra}). By noting that ${\cal 
T}_{hf}({\mathbf r})$ vanishes any time $x+y$ is odd, one obtains for 
$G^{-+}({\mathbf r})$ the following result
\begin{equation}
G^{-+}({\mathbf r})=(-)^{x+y}\{\r_{hf}^{4}\sum_{\G_{1}\G_{2}}w_{\G_{1}}w_{\G_{2}}
G^{[hf]}_{\G_{1}\G_{2}}({\mathbf r})^{2}+\frac{1}{N^{4}}{\cal T}({\mathbf r})
[{\cal T}({\mathbf r})+{\cal T}_{hf}({\mathbf r})]\}
\label{gmpss}
\end{equation}
where we have used Eqs.(\ref{ghfgg})(\ref{norm2})(\ref{t1dir}).

In order to make this result more explicit, we perform the sum in  
the first term of Eq.(\ref{gmpss}). It can be easily calculated coming 
back to the original $c$ picture. Indeed
\begin{equation}
(-)^{x+y}\sum_{\G_{1}\G_{2}}w_{\G_{1}}w_{\G_{2}}G^{[hf]}_{\G_{1}\G_{2}}({\mathbf r})^{2}=
\bra\Q^{[0]}_{AF}|c^{\dag}_{1,\da}c_{1,\ua}\hat{T}({\mathbf r})c^{\dag}_{1,\ua}c_{1,\da}
|\Q^{[0]}_{AF}\ket=
\bra\F^{[0]}_{\da}|c^{\dag}_{1,\da}c_{1,\ua}\hat{T}({\mathbf r})c^{\dag}_{1,\ua}c_{1,\da}
|\F^{[0]}_{\da}\ket\equiv X({\mathbf r})
\label{lastp}
\end{equation}
where $|\F^{[0]}_{\da}\ket$ is defined in Eq.(\ref{fizerobasso}) and 
$\hat{T}({\mathbf r})$ is the translation operator by ${\mathbf r}$: 
$\hat{T}^{\dag}({\mathbf r})c_{1}\hat{T}({\mathbf r})=c_{1}({\mathbf r})$. As usual $c_{i}$ is 
given by the same expression which defines $d_{i}$, with $d_{{\mathbf k}}\ra c_{{\mathbf k}}$. 
Therefore, according to the new expression of $d_{1}=d_{1}(0)$, we 
have $c_{1}=\frac{1}{|{\cal S}_{hf}|}\sum_{\e({\mathbf k})=0}c_{{\mathbf k}}$ and this is 
why  $c_{1}=c_{1}(0)=c_{1}({\mathbf r}=(0,0))$ shows up in Eq.(\ref{lastp}). 
Finally we observe that the spin-dependent filled Fermi 
sea $|\S_{\s}\ket$ can contribute 
only a phase factor corresponding to its momentum; since 
$|\S_{\s}\ket$ has vanishing momentum the phase factor is exactly 1.

The explicit evaluation of $X({\mathbf r})$ is deferred to Appendix 
\ref{xdierre}. Here we report the final result
\begin{equation}
\fbox{$
X({\mathbf r})=\frac{(-)^{x+y}}{|{\cal S}_{hf}|^{2}}\left\{
\begin{array}{ll}
    A+B\times {\cal T}^{2}_{hf}({\mathbf r}) & \;\;\;x+y\;{\mathrm even} \\
    A+B\times (4N-4) & \;\;\;x+y\;{\mathrm odd} \\
\end{array}\right. $}
\end{equation}
where $A$ and $B$ are two $N$-dependent constants. Eventually, 
substituing this last result in Eq.(\ref{gmpss}) and taking into 
account Eq.(\ref{gspindigmp}), we get the full 
analytic expression of the spin correlation function
\begin{equation}
\fbox{$ G_{\mathrm{spin}}({\mathbf r})=\frac{3}{2}\frac{(-)^{x+y}}{N^{4}}
\left[{\cal T}({\mathbf r})[{\cal T}({\mathbf r})+{\cal T}_{hf}({\mathbf r})]+
\left\{
\begin{array}{ll}
    A+B\times {\cal T}^{2}_{hf}({\mathbf r}) & \;\;\;x+y\;{\mathrm even} \\
    A+B\times (4N-4) & \;\;\;x+y\;{\mathrm odd} \\
\end{array}\right. \right] $}
\label{fullanags}
\end{equation}
In this form it is not hard to show that independent of the numerical 
value of the two constants $A$ and $B$ the sum rule 
\begin{equation}
    \sum_{{\mathbf r}}G_{\mathrm{spin}}({\mathbf r})=0
\label{srgs}
\end{equation}
holds. Indeed, let us consider the identities 
\begin{equation}
\sum_{{\mathbf r}}(-1)^{x+y}{\cal T}({\mathbf r})^{2}=
\sum_{{\mathbf r}}\sum_{\e({\mathbf k}),\e({\mathbf k}')<0}(-1)^{x+y}
e^{-i({\mathbf k}+{\mathbf k}')\cdot {\mathbf r}}=
\sum_{\e({\mathbf k}),\e({\mathbf k}')<0}\sum_{{\mathbf r}}
e^{-i({\mathbf k}+{\mathbf k}'+{\mathbf Q})\cdot {\mathbf r}};
\label{mare}
\end{equation}
since the ${\mathbf r}$ summation yields $N^{2}$ times a $\d$ 
function, and the $\d$ function is never satisfied (if $\e({\mathbf 
k})<0$ then $\e({\mathbf Q}-{\mathbf k})>0$), one finds that 
\begin{equation}
\sum_{{\mathbf r}}(-1)^{x+y}{\cal T}({\mathbf r})^{2}=0,
\label{mareantimare}
\end{equation}
and, similarly
\begin{equation}
\sum_{{\mathbf r}}(-1)^{x+y}{\cal T}({\mathbf r}){\cal T}_{hf}({\mathbf r})=0.
\label{marehf}
\end{equation}
On the other hand, since
\begin{equation}
\sum_{{\mathbf r }\in {\cal A}} e^{i{\mathbf k}\cdot {\mathbf r}}=
\frac{N^{2}}{2}(\d_{{\mathbf k},0}+\d_{{\mathbf k},{\mathbf Q}}),
\label{hfhf}
\end{equation}
where ${\cal A}$ is the sublattice with sites having $x+y$ even, one gets
\begin{equation}
\sum_{{\mathbf r}\in {\cal A}}{\cal T}_{hf}({\mathbf r})^{2}=
\frac{N^{2}}{2}(4N-4)
\label{trhf}
\end{equation}
and hence Eq.(\ref{srgs}). 

\section{The Charge Correlation Function}
\label{charge}

The charge and spin correlation functions are closely related. Let 
$\hat{n}_{{\mathbf r}}^{[\pm]}=\hat{n}_{{\mathbf r}\ua}
\pm\hat{n}_{{\mathbf r}\da}$; so, $\hat{n}_{{\mathbf r}}^{[+]}$
is the number operator, while $\hat{n}_{{\mathbf r}}^{[-]}$ 
is twice the $z$ component of the spin 
on the site ${\mathbf r}$. Then, the charge 
correlation function is given by 
\begin{equation}
G_{{\mathrm charge}}({\mathbf r})\equiv\bra\Q^{[0]}_{AF}|\hat{n}_{{\mathbf r}}^{[+]}
\hat{n}_{0}^{[+]}|\Q^{[0]}_{AF}\ket\;,
\label{fdccc}
\end{equation}
while the spin correlation function can be written as 
\begin{equation}
G_{{\mathrm spin}}({\mathbf r})=3
\bra\Q^{[0]}_{AF}|\hat{S}^{z}_{{\mathbf r}}\hat{S}^{z}_{0}|\Q^{[0]}_{AF}\ket=
\frac{3}{4}\bra\Q^{[0]}_{AF}|\hat{n}_{{\mathbf r}}^{[-]}
\hat{n}_{0}^{[-]}|\Q^{[0]}_{AF}\ket,
\label{fdcss}
\end{equation}
where $\hat{n}_{0}^{[\pm]}\equiv\hat{n}_{{\mathbf r}=(0,0)}^{[\pm]}$. Let 
\begin{equation}
|\Q^{[0]}_{AF}\ket=|\Q^{[0]}_{\ua}\ket-|\Q^{[0]}_{\da}\ket,
\end{equation}
with $|\Q^{[0]}_{\s}\ket=|\F^{[0]}_{\s}\ket\otimes|\S\ket$, be a 
suitable decomposition of the singlet ground state wave function, see 
Eq.(\ref{decosm}). Exploiting the 
invariance of $\bra\Q^{[0]}_{\s}|\hat{n}_{{\mathbf r}}^{[\pm]}
\hat{n}_{0}^{[\pm]}|\Q^{[0]}_{\s'}\ket$ 
for simultaneous flips of  $\s$ and $\s'$ we get
\begin{equation}
G_{{\mathrm charge}}({\mathbf r}) =
2\left[\bra\Q^{[0]}_{\ua}|\hat{n}_{0}^{[+]}\hat{T}({\mathbf r})
\hat{n}_{0}^{[+]}|\Q^{[0]}_{\ua}\ket-
\bra\Q^{[0]}_{\da}|\hat{n}_{0}^{[+]}\hat{T}({\mathbf r})
\hat{n}_{0}^{[+]}|\Q^{[0]}_{\ua}\ket\right]
\end{equation}
\begin{equation}
G_{{\mathrm spin}}({\mathbf r})=
\frac{3}{2}\left[\bra\Q^{[0]}_{\ua}|\hat{n}_{0}^{[-]}
\hat{T}({\mathbf r})\hat{n}_{0}^{[-]}|\Q^{[0]}_{\ua}\ket-
\bra\Q^{[0]}_{\da}|\hat{n}_{0}^{[-]}\hat{T}({\mathbf r})
\hat{n}_{0}^{[-]}|\Q^{[0]}_{\ua}\ket\right],
\end{equation}
where $\hat{T}({\mathbf r})$ is the operator of the translation  by ${\mathbf 
r}$, such that 
\begin{equation}
\hat{n}_{{\mathbf r}\s}=\hat{T}^{\dag}({\mathbf r})\hat{n}_{0\s}\hat{T}({\mathbf r})
\end{equation}  
and $\hat{T}({\mathbf r})|\Q^{[0]}_{AF}\ket=|\Q^{[0]}_{AF}\ket$ has been used. 
The action of $\hat{n}_{0}^{[\pm]}$ on the state 
$|\Q^{[0]}_{\s}\ket$ can be easily evaluated. We 
can express $c_{{\mathbf r}}$ as the sum of three operators like in Eq.(\ref{deco})
\begin{equation}
c_{{\mathbf r}}= \r_{hf}c_{1}({\mathbf r})+\r[c_{\x}({\mathbf r})+c_{\bar{\x}}({\mathbf r})]
\end{equation}
where $c_{1}$, $c_{\x}$ and $c_{\bar{\x}}$ are defined as $d_{1}$, 
$d_{\x}$ and $d_{\bar{\x}}$ in Eq.(\ref{dtra}), but $d_{{\mathbf k}}$ must be 
substituted with $c_{{\mathbf k}}$. Then we get
\begin{equation}
\hat{n}_{0}^{[\pm]}|\Q^{[0]}_{\ua}\ket=
(\hat{n}_{0\ua}\pm\hat{n}_{0\da})|\Q^{[0]}_{\ua}\ket=
\left[ 
(\r_{hf}^{2}+\r^{2}\pm \r^{2})+
\r\r_{hf}(c^{\dag}_{\bar{\x}\ua}c_{1,\ua}\pm c^{\dag}_{1,\da}
c_{\x\da})+
\r^{2}(c^{\dag}_{\bar{\x}\ua}c_{\x\ua}\pm
c^{\dag}_{\bar{\x}\da}c_{\x\da})
\right]|\Q^{[0]}_{\ua}\ket,
\label{adnpmsqa}
\end{equation}
\begin{equation}
\hat{n}_{0}^{[\pm]}|\Q^{[0]}_{\da}\ket=
(\hat{n}_{0\ua}\pm\hat{n}_{0\da})|\Q^{[0]}_{\da}\ket=
\left[
(\r^{2}\pm \r_{hf}^{2}\pm \r^{2})+
\r\r_{hf}( c^{\dag}_{1,\ua}
c_{\x\ua}\pm c^{\dag}_{\bar{\x}\da}c_{1,\da})+
\r^{2}(c^{\dag}_{\bar{\x}\ua}c_{\x\ua}\pm
c^{\dag}_{\bar{\x}\da}c_{\x\da})
\right]|\Q^{[0]}_{\da}\ket.
\label{adnpmsqb}
\end{equation}
Hence $(\hat{n}_{0\ua}\pm\hat{n}_{0\da})|\Q_{\s}\ket$ can be expressed 
as a linear combination of five orthogonal states. By means of these 
two last equations one gets
\begin{eqnarray}
    \bra\Q^{[0]}_{\ua}|\hat{n}_{0}^{[\pm]}
\hat{T}({\mathbf r})\hat{n}_{0}^{[\pm]}|\Q^{[0]}_{\ua}\ket=
[(\r^{2}_{hf}+\r^{2}\pm\r^{2})^{2}+2\r^{4}
\bra\S|c^{\dag}_{\x\ua}c_{\bar{\x}\ua}\hat{T}({\mathbf r})
c^{\dag}_{\bar{\x}\ua}c_{\x\ua}|\S\ket]
\bra\F^{[0]}_{\ua}|\hat{T}({\mathbf r})|\F^{[0]}_{\ua}\ket+
\nonumber \\ 
\r^{2}\r_{hf}^{2}(\bra\F^{[0]}_{\ua}|c^{\dag}_{1,\ua}
\hat{T}({\mathbf r})c_{1,\ua}|\F^{[0]}_{\ua}\ket
\bra\S|c_{\bar{\x}\ua}\hat{T}({\mathbf r})c^{\dag}_{\bar{\x}\ua}|\S\ket+
\bra\F^{[0]}_{\ua}|c_{1,\da}
\hat{T}({\mathbf r})c^{\dag}_{1,\da}|\F^{[0]}_{\ua}\ket
\bra\S|c^{\dag}_{\x\da}\hat{T}({\mathbf r})c_{\x\da}|\S\ket)
\label{ppsi}
\end{eqnarray}
and
\begin{equation}
\bra\Q^{[0]}_{\da}|\hat{n}_{0}^{[\pm]}
\hat{T}({\mathbf r})\hat{n}_{0}^{[\pm]}|\Q^{[0]}_{\ua}\ket=
[\pm(\r_{hf}^{2}+\r^{2}\pm \r^{2})^{2}+2\r^{4}
\bra\S|c^{\dag}_{\x\ua}c_{\bar{\x}\ua}\hat{T}({\mathbf r})
c^{\dag}_{\bar{\x}\ua}c_{\x\ua}|\S\ket]\bra\F^{[0]}_{\da}|\hat{T}({\mathbf r})|
\F^{[0]}_{\ua}\ket
\label{sppi}
\end{equation}
since $\bra\S|c^{\dag}_{\x\s}c_{\bar{\x}\s}\hat{T}({\mathbf r})
c^{\dag}_{\bar{\x}\s}c_{\x\s})|\S\ket$ does not depend on $\s$.  
The number of scalar products can be further reduced: we recall that 
$|\F_{AF}^{[0]}\ket\equiv|\F^{[0]}_{\ua}\ket-|\F^{[0]}_{\da}\ket$ 
is an eigenstate of the total momentum with 
vanishing eigenvalue:
\begin{equation}
1=\bra\F_{AF}^{[0]}|\hat{T}({\mathbf r})|\F_{AF}^{[0]}\ket=2\left[
\bra\F^{[0]}_{\ua}|\hat{T}({\mathbf r})|\F^{[0]}_{\ua}\ket-
\bra\F^{[0]}_{\da}|\hat{T}({\mathbf r})||\F^{[0]}_{\ua}\ket\right]
\end{equation}
and hence
\begin{equation}
\fbox{$\bra\F^{[0]}_{\da}|\hat{T}({\mathbf r})|\F^{[0]}_{\ua}\ket=
\bra\F^{[0]}_{\ua}|\hat{T}({\mathbf r})|\F^{[0]}_{\ua}\ket-\frac{1}{2}$}.
\label{aatab}
\end{equation}
Substituting  Eq.(\ref{aatab}) into Eq.(\ref{sppi}) and then 
subtracting   Eq.(\ref{ppsi}) term by term yields
\begin{eqnarray}
\bra\Q^{[0]}_{\ua}|\hat{n}_{0}^{[\pm]}\hat{T}({\mathbf r})
\hat{n}_{0}^{[\pm]}|\Q^{[0]}_{\ua}\ket-
\bra\Q^{[0]}_{\da}|\hat{n}_{0}^{[\pm]}\hat{T}({\mathbf r})
\hat{n}_{0}^{[\pm]}|\Q^{[0]}_{\ua}\ket=
\pm\frac{1}{2}(\r_{hf}^{2}+\r^{2}\pm \r^{2})^{2}+
\r^{4}\bra\S|c^{\dag}_{\x\ua}c_{\bar{\x}\ua}\hat{T}({\mathbf r})
c^{\dag}_{\bar{\x}\ua}c_{\x\ua}|\S\ket+
\nonumber \\ +
\r^{2}\r_{hf}^{2}\left[\bra\F^{[0]}_{\ua}|c^{\dag}_{1,\ua}\hat{T}({\mathbf r})
c_{1,\ua}|\F^{[0]}_{\ua}\ket
\bra\S|c_{\bar{\x}\ua}\hat{T}({\mathbf r})c^{\dag}_{\bar{\x}\ua}|\S\ket+
\bra\F^{[0]}_{\ua}|c_{1,\da}\hat{T}({\mathbf r})
c^{\dag}_{1,\da}|\F^{[0]}_{\ua}\ket
\bra\S|c^{\dag}_{\x\da}
\hat{T}({\mathbf r})c_{\x\da}|\S\ket\right]+\nonumber \\ +
\left[(1\pm(-1))(\r_{hf}^{2}+\r^{2}\pm \r^{2})^{2}\right]
\bra\F^{[0]}_{\ua}|\hat{T}({\mathbf r})|\F^{[0]}_{\ua}\ket
\label{ueftcascf}
\end{eqnarray}
and so, since $\r_{hf}^{2}+2\r^{2}=1$,
\begin{equation}
\fbox{$
G_{{\mathrm charge}}({\mathbf r})=1+\frac{4}{3}G_{{\mathrm spin}}({\mathbf r})+
\r_{hf}^{4}[1-4Y({\mathbf r})], $}
\end{equation}
where
\begin{equation}
    Y({\mathbf r})\equiv 
\bra\F^{[0]}_{\ua}|\hat{T}({\mathbf r})|
\F^{[0]}_{\ua}\ket.
\end{equation}
We postpone to Appendix \ref{ydierre} the explicit calculation of $Y({\mathbf r})$. 
Here we limit to present the final results
\begin{equation}
\fbox{$
Y({\mathbf r})=\frac{1}{4}+\frac{(-)^{x+y}}{|{\cal S}_{hf}|^{2}}\times 
\left\{\begin{array}{ll}
D+E\times {\cal T}_{hf}({\mathbf r})^{2} & x+y\;{\mathrm even} \\
D+E\times (4N-4) & x+y\;{\mathrm odd} \end{array}\right. $}
\label{fory}
\end{equation}
As for the spin correlation function one can easy verify that 
independent of the numerical value of the two $N$-constants $D$ and 
$E$, the sum rule 
\begin{equation}
\sum_{{\mathbf r}}G_{{\mathrm charge}}({\mathbf r})=N^{2}
\label{srgc}
\end{equation}
is satisfied.

\section{Results and Discussion}
\label{comp}

Most of the available data on the half filled Hubbard Model on a 
square lattice refer to the $4 \times 4$ cluster, see for example 
Ref.\cite{galan}. In the left hand side of 
Figure \ref{fre} we report a classical representation of the spin 
correlations in the $4 \times 4$ lattice: the length of the 
lines is proportional to the absolute 
value of the correlation function, and the sign is positive for the 
lines going up. This representation was adopted in Ref.\cite{galan} 
and our result is identical to that reported there, which was obtained  by second-order 
perturbation theory on the computer. 
\begin{figure}[H]
\begin{center}
	\epsfig{figure=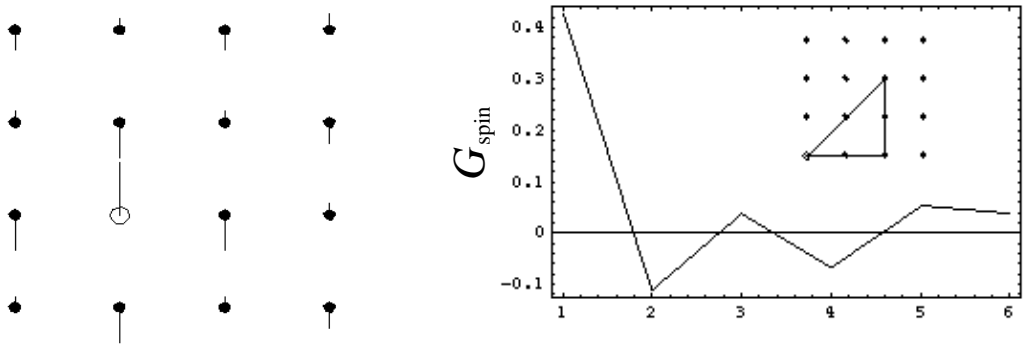,width=15cm}\caption{\footnotesize{
	Left. The spin correlation function between the origin (empty circle) and the 
	other sites; the length of the lines is proportional to the absolute 
	value of the correlation function, and the sign is positive for the 
	lines going up. The on-site value is reduced by a factor 0.4 for 
	graphical convenience. 
	Right. The spin correlation function in real space for the $4 \times 
	4$ model, along a counterclockwise path from the origin (empty 
	circle, see the inset). }}
\label{fre}
\end{center}
\end{figure}
More data\cite{fop}\cite{fop90}  on the $4 \times 4$  cluster 
were obtained by Fano, Ortolani and Parola by exact diagonalisation 
augmented by an intensive use of Group theory techniques. In 
the right hand side of Figure \ref{fre}  
we report the spin correlation function in real space,  
$G_{\mathrm{spin}}({\mathbf r})$ in Eq.(\ref{fullanags}), 
along a triangular path. Although our results are exact for $U 
\rightarrow 0$, remarkably the trend is quite the same as the one reported in 
Ref. \cite{fop} for $U=4$. 
An overall factor of 4 depends from the definition of the 
spin operators in \cite{fop}, lacking the usual $1/2$ factor. 

The analytic expression of the spin correlation function in 
Eq.(\ref{fullanags}) agree with  the important  Shen-Qiu-Tian
\cite{sqt} theorem, which has been extended to finite temperatures 
quite recently \cite{gst}. This theorem states that the spin correlation 
function must be positive on one sublattice and negative on the other. 
This applies to the results for the $4\times 4$, $6\times 6$, $8\times 
8$ and $10\times 10$ clusters as well, as shown 
in Figure \ref{spin16x16}. Essentially, the Shen-Qiu-Tian property is 
a consequence of the positive semidefinite ground state Lieb matrix, 
and we have explicitly verified this property in Section \ref{parth} above.
\begin{figure}[H]
\begin{center}
	\epsfig{figure=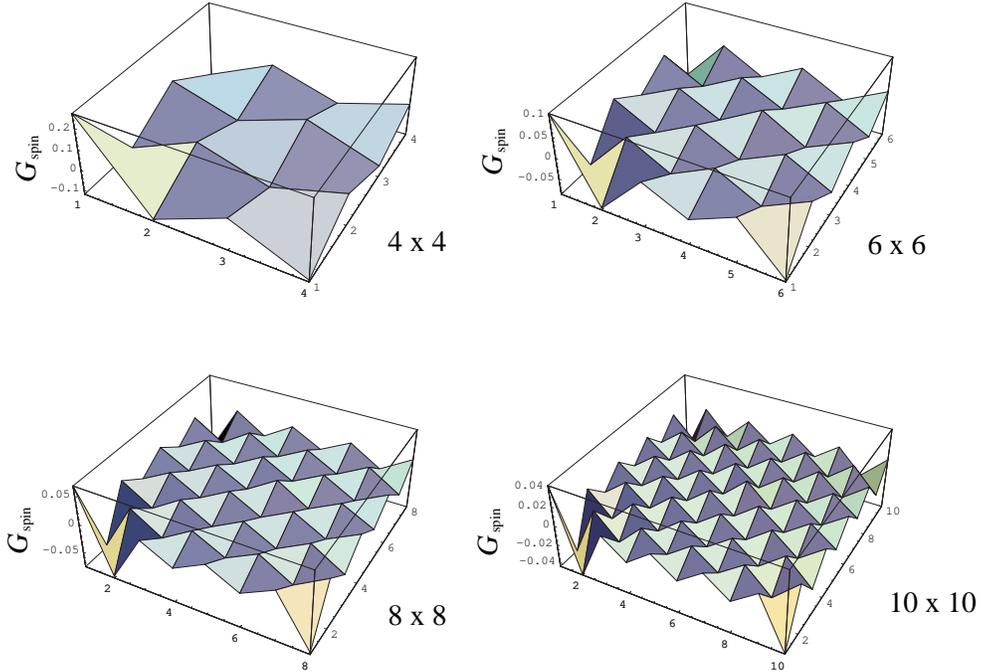,width=13cm}\caption{\footnotesize{
	The spin correlation function in real space for the 
	 $4\times 4$, $6\times 6$, $8\times 
         8$ and $10\times 10$
	 clusters. The Shen-Qiu-Tian property is evident.}}
\label{spin16x16}
\end{center}
\end{figure}
The Fourier transform of the spin correlation function 
\begin{equation}
G_{\mathrm{spin}}({\mathbf k})\equiv 
\sum_{{\mathbf r}}e^{i{\mathbf k}\cdot{\mathbf r}}
G_{\mathrm{spin}}({\mathbf r})
\end{equation}
is shown 
in Figure \ref{trasfospin}; the ticks on the $x$ axis correspond to  
the points $\G \equiv (0,0)$, $P \equiv (\p,0)$, $Q \equiv (\p,\p)$ 
and $\G$ again, in $k$-space. The trend is seen to converge rather quickly to a 
characteristic shape which is strongly peaked in the $Q$ direction.
\begin{figure}[H]
\begin{center}
	\epsfig{figure=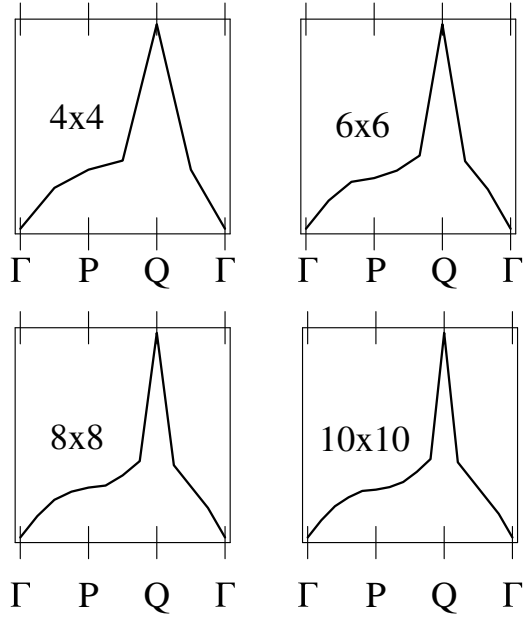,width=14cm}\caption{\footnotesize{
	The Fourier transform of the spin correlation functions in 
	clusters of varoius sizes. The ticks of the $x$ axis correspond to the 
	$\G$, $P$, $Q$ and $\G$, as usual. }}
\label{trasfospin}
\end{center} 
\end{figure}
The charge correlation function in real space shows characteristic 
structures with two intersecting channels at 45 degrees from the axes 
as exemplified in Figure \ref{charge3d} for the 10$\times$10 case; at 
the intersection the correlation function presents a narrow hole. 
Similar trends are observed for the other clusters, although the 
intensity of the corrugation declines with increasing cluster size.
\begin{figure}[H]
\begin{center}
	\epsfig{figure=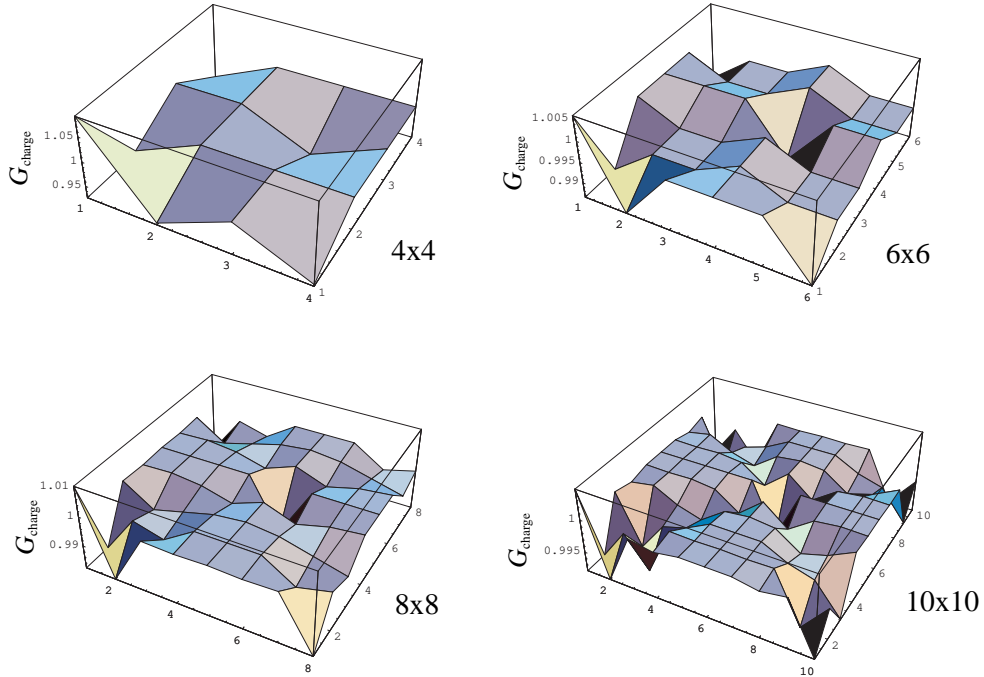,width=13cm}\caption{\footnotesize{
	Real space representation of the charge correlation function in 
	the $4\times 4$, $6\times 6$, $8\times 
         8$ and $10\times 10$
	clusters. 
	 }}
\label{charge3d}
\end{center}
\end{figure}

The Fourier transformed charge correlation function is dominated by a 
delta function at $\G \equiv (0,0)$ resulting from the almost 
constant distribution in real space. In 
Figure \ref{four10x10charge}  we have 
removed that delta. The figure represents 
$G_{{\mathrm charge}}({\mathbf k})\equiv \sum_{{\mathbf r}}
e^{i{\mathbf k}\cdot{\mathbf r}}G_{{\mathrm charge}}({\mathbf r})$ 
along the path $\G$, $P$, $Q$ and $\G$ (see figure caption for 
details) for the $10\times 10$ square lattice. Already at this cluster size 
$G_{{\mathrm charge}}({\mathbf k})$ shows a  very similar trend to its 
asymptotic ($N\ra\inf$) shape. 
\begin{figure}[H]
\begin{center}
	\epsfig{figure=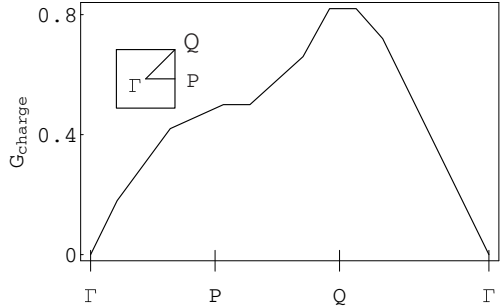,width=7cm}\caption{\footnotesize{
	The Fourier transform of the charge correlation function in 
	the 10 $\times$10
	cluster;  a delta function at the $\G$ point is removed. 
	The ticks of the $x$ axis correspond to the 
	$\G$, $P$, $Q$ and $\G$, as usual. }}
\label{four10x10charge}
\end{center}
\end{figure}

In conclusion, we have obtained  explicit analytic expressions for 
the spin and charge correlation functions using a weak-coupling 
ground state wave function $|\Q^{[0]}_{AF}\ket$ of the half filled Hubbard Model on 
a square lattice. We  compared our analytic results with the numerical 
data  available in the litterature. They always agree well; remarkably,
provided that $U\leq t/N^{2}$, our predictions are  good 
approximations to the exact diagonalization results. 

 As far as the non half-filled system is concerned, the same
local
formalism can be used to calculate the correlation functions of the
doped Hubbard antiferromagnet, but first-order perturbation theory is
not enough to single out a unique ground state. However, in the half filled 
ground state there are $2 N-2$ particles in the $ \epsilon =0$ shell that do 
not have double occupation; therefore,  doping
the system with two holes, one obtains a first-order ground
state provided that a pair is annihilated belonging to the shell ${\cal 
S}_{hf}$; this must be a $W=0$ pair.
Since
there are $W=0$ pairs belonging to different irreps of the Space Group,  the
 many-body ground state which is formed by annihilating the pair also has 
 components of different symmetries. 
For each symmetry, we 
shall have different correlation functions to compute. Actually, we need
second-order perturbation theory to resolve the degeneracy. This was
done in Ref.\cite{epj2001} in the special case $N=4$. On the
other hand, the problem becomes trivial when the shell at the Fermi surface is
totally filled since the non-interacting ground state is unique.

 As pointed out in Ref.\cite{mv},
one can use standard perturbation theory to calculate the correlation
functions order by order in $U$; in the thermodynamic limit they can be expanded
as  an  asymptotic series.

\appendix

\section{Evaluation of $X$}
\label{xdierre}

To evaluate $X({\mathbf r})=\bra\F^{[0]}_{\da}|c^{\dag}_{1,\da}c_{1,\ua}
\hat{T}({\mathbf r})c^{\dag}_{1,\ua}c_{1,\da}
|\F^{[0]}_{\da}\ket$ we need to  consider separately the cases of even 
and odd $x+y$. In the first case, let $T({\mathbf r})$ be the block-diagonal 
translation matrix in the local basis:
\begin{equation}
\hat{T}({\mathbf r})c^{\dag}_{i}|0\ket=\sum_{\g=1}^{2N-2}T({\mathbf 
r})_{i,\g}c^{\dag}_{\g}|0\ket=
\left\{\begin{array}{ll}
\sum_{\a=1}^{N-1}T({\mathbf r})_{i,\a}c^{\dag}_{\a}|0\ket, & \;\;i=1,\ldots,N-1 
\\ & \\
\sum_{\b=N}^{2N-2}T({\mathbf r})_{i,\b}c^{\dag}_{\b}|0\ket, & \;\;i=N,\ldots,2N-2
\end{array}\right. .
\label{htaocec}
\end{equation}
We have
\begin{eqnarray}
\hat{T}({\mathbf r})c^{\dag}_{1,\ua}c_{1,\da}|\F^{[0]}_{\da}\ket=
\frac{1}{\sqrt{{\cal N}}}
\sum_{\a_{1}..\a_{N-1}=1}^{N-1} \sum_{\b_{1}..\b_{N-1}=N}^{2N-2}
\prod_{a=1}^{N-1}T({\mathbf r})_{a,\a_{a}} 
\prod_{b=N}^{2N-2}T({\mathbf r})_{b,\b_{b}}
\times \nonumber \\ \times 
\sum_{k=0}^{N-2}(-)^{k}f_{k} 
\sum_{i_{k}>..>i_{1}=2}^{N-1} \sum_{j_{k}>..>j_{1}=1}^{N-1}
\hat{S}^{+}_{\a_{i_{k}}}..\hat{S}^{+}_{\a_{i_{1}}}
\hat{S}^{+}_{\a_{1}}
\hat{S}^{-}_{\b_{j_{k}}}..\hat{S}^{-}_{\b_{j_{1}}}
c^{\dag}_{\a_{1},\da}..c^{\dag}_{\a_{N-1},\da}
c^{\dag}_{\b_{1},\ua}..c^{\dag}_{\b_{N-1},\ua}|0\ket
\end{eqnarray}
and hence
\begin{eqnarray}
X({\mathbf r})=\frac{1}{{\cal N}}
\sum_{\a_{1}..\a_{N-1}=1}^{N-1} \sum_{\b_{1}..\b_{N-1}=N}^{2N-2}
\prod_{a=1}^{N-1}T({\mathbf r})_{a,\a_{a}} 
\prod_{b=N}^{2N-2}T({\mathbf r})_{b,\b_{b}}
\sum_{k=0}^{N-2}f^{2}_{k} 
\times \nonumber \\ \times
\{\sum_{m_{k}>..>m_{1}=2}^{N-1} \sum_{i_{k}>..>i_{1}=2}^{N-1}
\bra 0|c_{N-1,\da}..c_{1,\da}\hat{S}^{-}_{m_{k}}..\hat{S}^{-}_{m_{1}}
\hat{S}^{-}_{1}
\hat{S}^{+}_{\a_{i_{k}}}..\hat{S}^{+}_{\a_{i_{1}}}
\hat{S}^{+}_{\a_{1}}
c^{\dag}_{\a_{1},\da}..c^{\dag}_{\a_{N-1},\da}|0\ket\}
\times \nonumber \\ \times
\{\sum_{n_{k}>..>n_{1}=N}^{2N-2}\sum_{j_{k}>..>j_{1}=1}^{N-1}
\bra 0|c_{2N-2,\ua}..c_{N,\ua}
\hat{S}^{+}_{n_{k}}..\hat{S}^{+}_{n_{1}}
\hat{S}^{-}_{\b_{j_{k}}}..\hat{S}^{-}_{\b_{j_{1}}}
c^{\dag}_{\b_{1},\ua}..c^{\dag}_{\b_{N-1},\ua}|0\ket\}.
\label{icsfs}
\end{eqnarray}
Taking into account that the annihilation operators in the second and 
third row of the above equation are all different (in particular their 
indices are $1,\ldots,N-1$ in the second row and $N,\ldots,2N-2$ in 
the third row), a great simplification takes place:
\begin{equation}
\sum_{i_{k}>..>i_{1}=2}^{N-1}
\hat{S}^{+}_{\a_{i_{k}}}..\hat{S}^{+}_{\a_{i_{1}}}
\hat{S}^{+}_{\a_{1}}
c^{\dag}_{\a_{1},\da}..c^{\dag}_{\a_{N-1},\da}|0\ket=
\ve_{\a_{1}..\a_{N-1}}
\sum_{i_{k}>..>i_{1}=1,\neq \a_{1}}^{N-1}
\hat{S}^{+}_{i_{k}}..\hat{S}^{+}_{i_{1}}
\hat{S}^{+}_{\a_{1}}
c^{\dag}_{1,\da}..c^{\dag}_{N-1,\da}|0\ket+\ldots
\label{essei}
\end{equation}
\begin{equation}
\sum_{j_{k}>..>j_{1}=1}^{N-1}
\hat{S}^{-}_{\b_{j_{k}}}..\hat{S}^{-}_{\b_{j_{1}}}
c^{\dag}_{\b_{1},\ua}..c^{\dag}_{\b_{N-1},\ua}|0\ket=
\tilde{\ve}_{\b_{1}..\b_{N-1}}
\sum_{j_{k}>..>j_{1}=N}^{2N-2}
\hat{S}^{-}_{j_{k}}..\hat{S}^{-}_{j_{1}}
c^{\dag}_{N,\ua}..c^{\dag}_{2N-2,\ua}|0\ket+\ldots
\label{essej}
\end{equation}
where $\ve$ is the totally antisymmetric tensor with $N-1$ indices, 
while $\tilde{\ve}_{\b_{1}..\b_{N-1}}\equiv
\ve_{\b_{1}-N+1..\b_{N-1}-N+1}$ and the dots mean that we are 
neglecting other terms whose contribution to the scalar product is zero. 
On the right hand side of Eq.(\ref{essei})  the summation indices 
$i_{k}>..>i_{1}$ run in the interval $\{1,\ldots,N-1\}$ in such a way 
that none of them is equal to $\a_{1}$. 
Using Eq.(\ref{essei}), the second row of Eq.(\ref{icsfs}) yields 
\begin{eqnarray}
\ve_{\a_{1}..\a_{N-1}}
\sum_{m_{k}>..>m_{1}=2}^{N-1} 
\sum_{i_{k}>..>i_{1}=1,\neq \a_{1}}^{N-1}
\bra 0|c_{N-1,\da}..c_{1,\da}\hat{S}^{-}_{m_{k}}..\hat{S}^{-}_{m_{1}}
\hat{S}^{-}_{1}
\hat{S}^{+}_{i_{k}}..\hat{S}^{+}_{i_{1}}
\hat{S}^{+}_{\a_{1}}
c^{\dag}_{1,\da}..c^{\dag}_{N-1,\da}|0\ket=\nonumber \\ =
\ve_{\a_{1}..\a_{N-1}}\left[
\d_{1\a_{1}}\left(\begin{array}{c} N-2 \\ k \end{array}\right)
+(1-\d_{1\a_{1}})
\left(\begin{array}{c} N-3 \\ k-1 \end{array}\right)\right]
\end{eqnarray}
while using Eq.(\ref{essej}), the third row of Eq.(\ref{icsfs}) yields
\begin{eqnarray}
\tilde{\ve}_{\b_{1}..\b_{N-1}}
\sum_{n_{k}>..>n_{1}=N}^{2N-2} \sum_{j_{k}>..>j_{1}=N}^{2N-2}
\bra 0|c_{2N-2,\ua}..c_{N,\ua}
\hat{S}^{+}_{n_{k}}..\hat{S}^{+}_{n_{1}}
\hat{S}^{-}_{j_{k}}..\hat{S}^{-}_{j_{1}}
c^{\dag}_{N,\ua}..c^{\dag}_{2N-2,\ua}|0\ket=
\nonumber \\ =
\tilde{\ve}_{\b_{1}..\b_{N-1}}
\left(\begin{array}{c} N-1 \\ k \end{array}\right).
\end{eqnarray}
These two last results allow to rewrite $X({\mathbf r})$ as 
\begin{equation}
X({\mathbf r})=\frac{A}{|{\cal S}_{hf}|^{2}}
+B{\cal C}_{1,1}({\mathbf r})T({\mathbf r})_{1,1}
\end{equation}
where, using the convention 
$\left(\begin{array}{c} r \\ -|s| \end{array}\right)=0$ for a binomial 
coefficient with negative down entry, 
\begin{equation}
    \fbox{$
A=\frac{|{\cal S}_{hf}|^{2}}{{\cal N}}
\sum_{k=0}^{N-2}f_{k}^{2}
\left(\begin{array}{c} N-1 \\ k \end{array}\right)
\left(\begin{array}{c} N-3 \\ k-1 \end{array}\right)$}
\end{equation}
\begin{equation}
    \fbox{$
B=\frac{1}{{\cal N}}
\sum_{k=0}^{N-2}f^{2}_{k}
\left(\begin{array}{c} N-1 \\ k \end{array}\right)\left[
\left(\begin{array}{c} N-2 \\ k \end{array}\right)-
\left(\begin{array}{c} N-3 \\ k-1 
\end{array}\right)\right]$}
\end{equation}
and ${\cal C}_{1,1}({\mathbf r})$ is the (1,1) algebraic complement of the matrix 
$T({\mathbf r})$ whose determinant is equal to one for even $x+y$. The (1,1) algebraic 
complement can be expressed in terms of the (1,1) element of the 
matrix $T({\mathbf r})$: 
${\cal C}_{1,1}({\mathbf r})=T^{\dag}({\mathbf r})_{1,1}
{\mathrm Det}[T({\mathbf r})]=T({\mathbf r})_{1,1}$ 
(since $T({\mathbf r})_{i,j}\in\Re$). Next one has to 
recognize that $T({\mathbf r})_{1,1}$ 
is, by definition, equal to $t_{1}({\mathbf r})$ (see Eq.(\ref{dunotra})) whose 
analytic expression is given in Eq.(\ref{t1dir}). Therefore, any time 
$x+y$ is even we have
\begin{equation}
\fbox{$
X({\mathbf r})=\frac{1}{|{\cal S}_{hf}|^{2}}[A
+B\times {\cal T}_{hf}({\mathbf r})^{2}]$}
\end{equation}
    
On the other hand, for odd $x+y$ the translation matrix in the local 
basis is antiblock diagonal, that is 
\begin{equation}
\hat{T}({\mathbf r})c^{\dag}_{i}|0\ket=
\sum_{\g=1}^{2N-2}T({\mathbf r})_{i,\g}c^{\dag}_{\g}|0\ket=
\left\{\begin{array}{ll}
\sum_{\a=N}^{2N-2}T({\mathbf r})_{i,\a}c^{\dag}_{\a}|0\ket, & \;\;i=1,\ldots,N-1 
\\ & \\
\sum_{\b=1}^{N-1}T({\mathbf r})_{i,\b}c^{\dag}_{\b}|0\ket, & \;\;i=N,\ldots,2N-2
\end{array}\right. 
\label{htaococ}
\end{equation}
and hence
\begin{eqnarray}
\hat{T}({\mathbf r})c^{\dag}_{1,\ua}c_{1,\da}|\F^{[0]}_{\da}\ket=
\frac{1}{\sqrt{{\cal N}}}
\sum_{\a_{1}..\a_{N-1}=N}^{2N-2} \sum_{\b_{1}..\b_{N-1}=1}^{N-1}
\prod_{a=1}^{N-1}T({\mathbf r})_{a,\a_{a}} 
\prod_{b=N}^{2N-2}T({\mathbf r})_{b,\b_{b}}
\times \nonumber \\ \times 
\sum_{k=0}^{N-2}(-)^{k}f_{k} 
\sum_{i_{k}>..>i_{1}=2}^{N-1} \sum_{j_{k}>..>j_{1}=1}^{N-1}
\hat{S}^{+}_{\a_{i_{k}}}..\hat{S}^{+}_{\a_{i_{1}}}
\hat{S}^{+}_{\a_{1}}
\hat{S}^{-}_{\b_{j_{k}}}..\hat{S}^{-}_{\b_{j_{1}}}
c^{\dag}_{\a_{1},\da}..c^{\dag}_{\a_{N-1},\da}
c^{\dag}_{\b_{1},\ua}..c^{\dag}_{\b_{N-1},\ua}|0\ket.
\end{eqnarray}
Let us consider the $k$-th term of this state. One can check by 
direct inspection that the only non vanishing contribution in the scalar 
product with the state $c^{\dag}_{1,\ua}c_{1,\da}|\F^{[0]}_{\da}\ket$ 
comes from the $(N-k-2)$-th term of the corresponding expansion, see 
Eq.(\ref{fizerobasso}). Hence
\begin{eqnarray}
X({\mathbf r})=-\frac{1}{{\cal N}}
\sum_{\a_{1}..\a_{N-1}=N}^{2N-2} \sum_{\b_{1}..\b_{N-1}=1}^{N-1}
\prod_{a=1}^{N-1}T({\mathbf r})_{a,\a_{a}} 
\prod_{b=N}^{2N-2}T({\mathbf r})_{b,\b_{b}}
\sum_{k=0}^{N-2}f_{k}f_{N-k-2}
\times \nonumber \\ \times
\{\sum_{m_{N-k-2}>..>m_{1}=2}^{N-1}
\sum_{j_{k}>..>j_{1}=1}^{N-1}
\bra 0|c_{N-1,\da}..c_{1,\da}
\hat{S}^{-}_{m_{N-k-2}}..\hat{S}^{-}_{m_{1}}\hat{S}^{-}_{1}
\hat{S}^{-}_{\b_{j_{k}}}..\hat{S}^{-}_{\b_{j_{1}}}
c^{\dag}_{\b_{1},\ua}..c^{\dag}_{\b_{N-1},\ua}|0\ket\}
\times \nonumber \\ \times
\{\sum_{n_{N-k-2}>..>n_{1}=N}^{2N-2}
\sum_{i_{k}>..>i_{1}=2}^{N-1}
\bra 0|c_{2N-2,\ua}..c_{N,\ua}
\hat{S}^{+}_{n_{N-k-2}}..\hat{S}^{+}_{n_{1}}
\hat{S}^{+}_{\a_{i_{k}}}..\hat{S}^{+}_{\a_{i_{1}}}
\hat{S}^{+}_{\a_{1}}
c^{\dag}_{\a_{1},\da}..c^{\dag}_{\a_{N-1},\da}|0\ket\}.
\label{icsoddfs}
\end{eqnarray}
Analogously to the case $x+y$ even, the particular structure of 
$|\F^{[0]}_{AF}\ket$ allows the following simplification
\begin{equation}
\sum_{j_{k}>..>j_{1}=1}^{N-1}
\hat{S}^{-}_{\b_{j_{k}}}..\hat{S}^{-}_{\b_{j_{1}}}
c^{\dag}_{\b_{1},\ua}..c^{\dag}_{\b_{N-1},\ua}|0\ket=
\ve_{\b_{1}..\b_{N-1}}\sum_{j_{k}>..>j_{1}=1}^{N-1}
\hat{S}^{-}_{j_{k}}..\hat{S}^{-}_{j_{1}}
c^{\dag}_{1,\ua}..c^{\dag}_{N-1,\ua}|0\ket+\ldots
\label{gsuno}
\end{equation}
\begin{equation}
\sum_{i_{k}>..>i_{1}=2}^{N-1}
\hat{S}^{+}_{\a_{i_{k}}}..\hat{S}^{+}_{\a_{i_{1}}}
\hat{S}^{+}_{\a_{1}}
c^{\dag}_{\a_{1},\da}..c^{\dag}_{\a_{N-1},\da}|0\ket=
\tilde{\ve}_{\a_{1}..\a_{N-1}}
\sum_{i_{k}>..>i_{1}=N,\neq \a_{1}}^{2N-2}
\hat{S}^{+}_{i_{k}}..\hat{S}^{+}_{i_{1}}
\hat{S}^{+}_{\a_{1}}
c^{\dag}_{N,\da}..c^{\dag}_{2N-2,\da}|0\ket\ldots
\label{gsdue}
\end{equation}
where 
$\tilde{\ve}_{\a_{1}..\a_{N-1}}=\ve_{\a_{1}-N+1\ldots\a_{N-1}-N+1}$ 
and the sum on the right hand side of Eq.(\ref{gsdue}) means that the 
indices $i_{k}>..>i_{1}$ run in the interval $\{N,\ldots,2N-2\}$ in 
such a way that no of them is equal to $\a_{1}$. All the neglected terms 
contribute nothing to the scalar product. 

Using Eq.(\ref{gsuno}) the second row of Eq.(\ref{icsoddfs}) yields
\begin{eqnarray}
\ve_{\b_{1}..\b_{N-1}}
\sum_{m_{N-k-2}>..>m_{1}=2}^{N-1}\sum_{j_{k}>..>j_{1}=1}^{N-1}
\bra 0|c_{N-1,\da}..c_{1,\da}
\hat{S}^{-}_{m_{N-k-2}}..\hat{S}^{-}_{m_{1}}\hat{S}^{-}_{1}
\hat{S}^{-}_{j_{k}}..\hat{S}^{-}_{j_{1}}
c^{\dag}_{1,\ua}..c^{\dag}_{N-1,\ua}|0\ket=
\nonumber \\=
\ve_{\b_{1}..\b_{N-1}}
\left(\begin{array}{c} N-2 \\ k \end{array}\right)
\end{eqnarray}
while using Eq.(\ref{gsdue}) the third row of Eq.(\ref{icsoddfs}) yields
\begin{eqnarray}
\tilde{\ve}_{\a_{1}..\a_{N-1}}
\sum_{n_{N-k-2}>..>n_{1}=N}^{2N-2}
\sum_{i_{k}>..>i_{1}=N,\neq \a_{1}}^{2N-2}
\bra 0|c_{2N-2,\ua}..c_{N,\ua}
\hat{S}^{+}_{n_{N-k-2}}..\hat{S}^{+}_{n_{1}}
\hat{S}^{+}_{i_{k}}..\hat{S}^{+}_{i_{1}}
\hat{S}^{+}_{\a_{1}}c^{\dag}_{N,\da}..c^{\dag}_{2N-2,\da}|0\ket=
\nonumber \\=
\tilde{\ve}_{\a_{1}..\a_{N-1}}
\left(\begin{array}{c} N-2 \\ k \end{array}\right).
\end{eqnarray}
Substituing these results in the expression for $X({\mathbf r})$ we get
\begin{equation}
    \fbox{$
    X({\mathbf r})=-\frac{C}{|{\cal S}_{hf}|^{2}}=-\frac{1}{|{\cal S}_{hf}|^{2}}
    [A+B\times (4N-4)]$}
    \label{oddicsfine}
\end{equation}
where we have taken into account that ${\mathrm Det}[T({\mathbf r})]=-1$ for odd 
$x+y$ and the constant $C$ is given by
\begin{equation}
    \fbox{$
C=\frac{|{\cal S}_{hf}|^{2}}{{\cal N}}
\sum_{k=0}^{N-2}f_{k}f_{N-k-2}
\left(\begin{array}{c} N-2 \\ k \end{array}\right)^{2}$}\;\;.
\end{equation}
In the last equality of Eq.(\ref{oddicsfine}) we have used 
$C=A+B\times (4N-4)$ which is a direct consequence of the sum 
rule for the spin correlation function, see Eq.(\ref{srgs}).

\section{Evaluation of $Y$}
\label{ydierre}

Here we show that $Y({\mathbf r})\equiv 
\bra\F^{[0]}_{\ua}|\hat{T}({\mathbf r})|
\F^{[0]}_{\ua}\ket$ has the form shown in Eq.(\ref{fory}) and we derive the 
explicit values for the two constants $D$ and $E$. As for $X({\mathbf r})$ we 
will first consider the case $x+y$ even and thereafter the case $x+y$ 
odd. Making use of Eq.(\ref{htaocec}), we get
\begin{eqnarray}
\hat{T}({\mathbf r})|\F^{[0]}_{\ua}\ket=\frac{1}{\sqrt{{\cal N}}}
\sum_{\a_{1}..\a_{N-1}=1}^{N-1}\sum_{\b_{1}..\b_{N-1}=N}^{2N-2}
\prod_{a=1}^{N-1}T({\mathbf r})_{a,\a_{a}}
\prod_{b=N}^{2N-2}T({\mathbf r})_{b,\b_{b}}
\times \nonumber \\ \times 
\sum_{k=0}^{N-2}(-1)^{k}f_{k}
\sum_{i_{k}>\ldots>i_{1}=2}^{N-1}
\sum_{j_{k}>\ldots>j_{1}=1}^{N-1}
\hat{S}^{-}_{\a_{i_{k}}}..\hat{S}^{-}_{\a_{i_{1}}}
\hat{S}^{+}_{\b_{j_{k}}}..\hat{S}^{+}_{\b_{j_{1}}}
c^{\dag}_{\a_{1},\ua}..c^{\dag}_{\a_{N-1},\ua}
c^{\dag}_{\b_{N},\da}..c^{\dag}_{\b_{2N-2},\da}|0\ket
\label{tdfacp}
\end{eqnarray}
The $k$-th term in the sum of Eq.(\ref{tdfacp}) gives 
non-vanishing scalar product only with the $k$-th term in 
Eq.(\ref{fizeroalto}) and hence 
\begin{eqnarray}
Y({\mathbf r})=
\frac{1}{{\cal N}}
\sum_{\a_{1}..\a_{N-1}=1}^{N-1}\sum_{\b_{1}..\b_{N-1}=N}^{2N-2}
\prod_{a=1}^{N-1}T({\mathbf r})_{a,\a_{a}}
\prod_{b=N}^{2N-2}T({\mathbf r})_{b,\b_{b}}\sum_{k=0}^{N-2}f_{k}^{2}
\times \nonumber \\  \times 
\{
\sum_{m_{k}>..>m_{1}=2}^{N-1}\sum_{i_{k}>..>i_{1}=2}^{N-1}
\bra 0|c_{N-1,\ua}..c_{1,\ua}
\hat{S}^{+}_{m_{k}}..\hat{S}^{+}_{m_{1}}
\hat{S}^{-}_{\a_{i_{k}}}..\hat{S}^{-}_{\a_{i_{1}}}
c^{\dag}_{\a_{1},\ua}..c^{\dag}_{\a_{N-1},\ua}|0\ket\}
\times\nonumber \\ \times 
\{
\sum_{n_{k}>..>n_{1}=N}^{2N-2}\sum_{j_{k}>..>j_{1}=1}^{N-1}
\bra 0|c_{2N-2,\da}..c_{N,\da}
\hat{S}^{-}_{n_{k}}..\hat{S}^{-}_{n_{1}}
\hat{S}^{+}_{\b_{j_{k}}}..\hat{S}^{+}_{\b_{j_{1}}}
c^{\dag}_{\b_{1},\da}..c^{\dag}_{\b_{N-1},\da}|0\ket\} .
\label{ssp}
\end{eqnarray}
Now we use the fact that the indices $\a_{1},\ldots,\a_{N-1}$ must be all 
different and within the range $\{1,\ldots,N-1\}$ otherwise the scalar 
product vanishes. This means that  
\begin{equation}
\sum_{i_{k}>..>i_{1}=2}^{N-1}\hat{S}^{-}_{\a_{i_{k}}}.. 
\hat{S}^{-}_{\a_{i_{1}}}
c^{\dag}_{\a_{1},\ua}..c^{\dag}_{\a_{N-1},\ua}|0\ket=
\ve_{\a_{1}..\a_{N-1}} 
\sum_{i_{k}>..>i_{1}=1,\neq \a_{1}}^{N-1}
\hat{S}^{-}_{i_{1}}.. \hat{S}^{-}_{i_{k}}
c^{\dag}_{1,\ua}..c^{\dag}_{N-1,\ua}|0\ket+\ldots
\end{equation}
where the neglected terms do not contribute to the scalar product. 
In the second row of Eq.(\ref{ssp}), 
$c_{1,\ua}$ commutes with all the raising spin operators whatever are 
the values of the $k$ indices $m_{k}>..>m_{1}$ in the range specified 
by the sum. 
This implies that $i_{1}$ cannot be 1. 
On the other hand $c^{\dag}_{\a_{1}}$ commutes with all the lowering 
spin operators and hence no one of the indices $m_{k}>\ldots>m_{1}$ 
can be $\a_{1}$ otherwise the corresponding term vanishes. 
Hence the term in the secon row gives 
\begin{eqnarray}
\sum_{m_{k}>..>m_{1}=2,\neq \a_{1}}^{N-1}
\sum_{i_{k}>..>i_{1}=2,\neq \a_{1}}^{N-1}
\ve_{\a_{1}..\a_{N-1}}\bra 0|c_{N-1,\ua}..c_{1,\ua}
\hat{S}^{+}_{m_{k}}..\hat{S}^{+}_{m_{1}}
\hat{S}^{-}_{i_{1}}.. \hat{S}^{-}_{i_{k}}
c^{\dag}_{1,\ua}..c^{\dag}_{N-1,\ua}|0\ket=
\nonumber \\ =
[\sum_{i_{k}>\ldots>i_{1}=2,\neq \a_{1}}^{N-1}]
\ve_{\a_{1}..\a_{N-1}}.
\end{eqnarray}
If $k=N-2$ the sum is zero except for $\a_{1}=1$ since the indices 
$i_{k}>\ldots>i_{1}$ don't have space to run. Hence 
\begin{eqnarray}
\sum_{i_{k}>..>i_{1}=2,\neq \a_{1}}^{N-1}=
\d_{k,N-2}\d_{1,\a_{1}}+(1-\d_{k,N-2})\left[
\left(\begin{array}{c}N-3\\ 
k\end{array}\right)+\d_{1,\a_{1}}(\left(\begin{array}{c}N-2\\ 
k\end{array}\right)-\left(\begin{array}{c}N-3\\ 
k\end{array}\right))\right]=\nonumber \\ =
(1-\d_{k,N-2})
\left(\begin{array}{c}N-3\\ 
k\end{array}\right)+\d_{1,\a_{1}}\left[
\d_{k,N-2}+(1-\d_{k,N-2})(\left(\begin{array}{c}N-2\\ 
k\end{array}\right)-\left(\begin{array}{c}N-3\\ 
k\end{array}\right))\right]
\end{eqnarray}
while the third row of Eq.(\ref{ssp}) yields
\begin{eqnarray}
\tilde{\ve}_{\b_{1}..\b_{N-1}}
\sum_{n_{k}>..>n_{1}=N}^{2N-2}\sum_{j_{k}>..>j_{1}=N}^{2N-2}
\bra 0|c_{2N-2,\da}..c_{N,\da}
\hat{S}^{-}_{n_{k}}..\hat{S}^{-}_{n_{1}}
\hat{S}^{+}_{j_{k}}..\hat{S}^{+}_{j_{1}}
c^{\dag}_{N,\da}..c^{\dag}_{2N-2,\da}|0\ket=
\nonumber \\ =
\tilde{\ve}_{\b_{1}..\b_{N-1}}
\left(\begin{array}{c}N-1\\ k\end{array}\right)
\end{eqnarray}
as can be verified by using the total antisymmetry of each 
homogeneous polynomial in the raising spin operators. 

Therefore for even $x+y$ one can write
\begin{equation}
\fbox{$
Y({\mathbf r})=\frac{1}{4}+
\frac{D}{|{\cal S}_{hf}|^{2}}+E
{\cal C}_{1,1}({\mathbf r})T({\mathbf r})_{1,1}=\frac{1}{4}+
\frac{1}{|{\cal S}_{hf}|^{2}}
[D+E\times {\cal T}_{hf}({\mathbf r})^{2}]$}\;\;,
\label{evotfeixiy}
\end{equation}
where $D$ and $E$ are two $N$-dependent constants given by:
\begin{equation}
\fbox{$
D=\frac{|{\cal S}_{hf}|^{2}}{{\cal N}}\left[
\sum_{k=0}^{N-3}f_{k}^{2}\left(\begin{array}{c}N-1\\ k\end{array}\right)
\left(\begin{array}{c}N-3\\ k\end{array}\right)-\frac{{\cal N}}{4}\right]
$}
\label{bn}
\end{equation}
\begin{equation}
\fbox{$
E=\frac{1}{{\cal N}}\left\{(N-1)f^{2}_{N-2}+
\sum_{k=1}^{N-3}f^{2}_{k}\left(\begin{array}{c}N-1\\ k\end{array}\right)\left[
\left(\begin{array}{c}N-2\\ k\end{array}\right)-
\left(\begin{array}{c}N-3\\ k\end{array}\right)\right]\right\}$}\;\;.
\label{cn}
\end{equation}

For odd $x+y$ we make use of Eq.(\ref{htaococ}). Then, 
the action of $\hat{T}({\mathbf r})$ over $|\F^{[0]}_{\ua}\ket$ gives 
\begin{eqnarray}
\hat{T}({\mathbf r})|\F^{[0]}_{\ua}\ket=\frac{1}{\sqrt{{\cal N}}}
\sum_{\a_{1}..\a_{N-1}=N}^{2N-2}\sum_{\b_{1}..\b_{N-1}=1}^{N-1}
\prod_{a=1}^{N-1}T({\mathbf r})_{a,\a_{a}}
\prod_{b=N}^{2N-2}T({\mathbf r})_{b,\b_{b}} 
\times \nonumber \\  \times 
\sum_{k=0}^{N-2}(-)^{k}f_{k}
\sum_{i_{k}>\ldots>i_{1}=2}^{N-1}\sum_{j_{k}>\ldots>j_{1}=1}^{N-1}
\hat{S}^{-}_{\a_{i_{k}}}..\hat{S}^{-}_{\a_{i_{1}}}
\hat{S}^{+}_{\b_{j_{k}}}..\hat{S}^{+}_{\b_{j_{1}}}
c^{\dag}_{\a_{1},\ua}..c^{\dag}_{\a_{N-1},\ua}
c^{\dag}_{\b_{1},\da}..c^{\dag}_{\b_{N-1},\da}|0\ket .
\label{aotofa}
\end{eqnarray}
In the scalar product with the $|\F^{[0]}_{\ua}\ket$ state, only the terms with the 
same number of up (or down) spins in the first $N-1$ (and hence in the 
last $N-1$) local states will survive. 
Let us consider for example the $k$-th term of the sum in the second  
row of Eq.(\ref{aotofa}); it contains states where $k$ of the 
last $N-1$ local states have spin down and $k$ of the first $N-1$ 
local states have spin up. 
This term has non-vanishing scalar product only with the $(N-k-1)$-th 
term of the sum in the definition of $|\F^{[0]}_{\ua}\ket$, 
Eq.(\ref{fizeroalto}). In 
particular this implies that the terms where the first and the last 
$N-1$ local states have all the spins aligned do not contribute to the 
scalar product. Hence 
\begin{eqnarray}
Y({\mathbf r})=\frac{1}{{\cal N}}
\sum_{\a_{1}..\a_{N-1}=N}^{2N-2}\sum_{\b_{1}..\b_{N-1}=1}^{N-1}
\prod_{a=1}^{N-1}T({\mathbf r})_{a,\a_{a}}
\prod_{b=N}^{2N-2}T({\mathbf r})_{b,\b_{b}}
\sum_{k=1}^{N-2}f_{k}f_{N-k-1}
\times \nonumber \\ \times 
\{
\sum_{m_{N-k-1}>..>m_{1}=2}^{N-1}\sum_{j_{k}>..>j_{1}=1}^{N-1}
\bra 0|c_{N-1,\ua}..c_{1,\ua}
\hat{S}^{+}_{m_{N-k-1}}..\hat{S}^{+}_{m_{1}}
\hat{S}^{+}_{\b_{j_{k}}}..\hat{S}^{+}_{\b_{j_{1}}}
c^{\dag}_{\b_{1}\da}..c^{\dag}_{\b_{N-1}\da}|0\ket\}
\times \nonumber \\ \times
\{
\sum_{n_{N-k-1}>..>n_{1}=N}^{2N-2}\sum_{i_{k}>..>i_{1}=2}^{N-1}
\bra 0|c_{2N-2,\da}..c_{N,\da}
\hat{S}^{-}_{n_{N-k-1}}..\hat{S}^{-}_{n_{1}}
\hat{S}^{-}_{\a_{i_{k}}}..\hat{S}^{-}_{\a_{i_{1}}}
c^{\dag}_{\a_{1},\ua}..c^{\dag}_{\a_{N-1},\ua}|0\ket\}.
\label{fsp}
\end{eqnarray}
Let us consider the term in the second row.  
For a given choice of $\b_{1}..\b_{N-1}$ one finds 
\begin{equation}
\sum_{j_{k}>..>j_{1}=1}^{N-1}
\hat{S}^{+}_{\b_{j_{k}}}..\hat{S}^{+}_{\b_{j_{1}}}
c^{\dag}_{\b_{1}\da}..c^{\dag}_{\b_{N-1}\da}|0\ket=
\ve_{\b_{1}..\b_{N-1}}\sum_{j_{k}>..>j_{1}=1}^{N-1}
\hat{S}^{+}_{j_{k}}.. 
\hat{S}^{+}_{j_{1}}c^{\dag}_{1,\da}..c^{\dag}_{N-1,\da}|0\ket+\ldots ,
\end{equation}
where the missed terms do not contribute to the scalar product. 
Since the $c_{1,\ua}$ annihilation operator commutes with all the 
raising spin operators coming from the sum over $m_{N-k-1}>..>m_{1}$,  
$j_{1}$ is constrained to be 1 for all non vanishing contributions. 
Still for $j_{k}>..>j_{2}$ fixed there is only one choice for 
$m_{N-k-1}>..>m_{1}$ to have non-vanishing result. In particular,  
the possible results for a given choice of $j_{k}>..>j_{2}$ and 
$m_{N-k-1}>..>m_{1}$ are 0 or 1. Hence the term in the first square bracket 
yields 
\begin{eqnarray}
\ve_{\b_{1}..\b_{N-1}}
\sum_{m_{N-k-1}>..>m_{1}=2}^{N-1}\sum_{j_{k}>..>j_{2}=2}^{N-1}
\bra 0|c_{N-1,\ua}..c_{1,\ua}
\hat{S}^{+}_{m_{N-k-1}}..\hat{S}^{+}_{m_{1}}
\hat{S}^{+}_{j_{k}}.. 
\hat{S}^{+}_{j_{1}}c^{\dag}_{1,\da}..c^{\dag}_{N-1,\da}|0\ket
=\nonumber \\ =\ve_{q_{1}..q_{N-1}}
\left(\begin{array}{c} N-2 \\ k-1 \end{array}\right) .
\end{eqnarray}
A similar trick can be used for the term in the third row  
of Eq.(\ref{fsp}). We get 
\begin{eqnarray}
\sum_{i_{k}>..>i_{1}=2}^{N-1}
\hat{S}^{-}_{\a_{i_{k}}}..\hat{S}^{-}_{\a_{i_{1}}}
c^{\dag}_{\a_{1},\ua}..c^{\dag}_{\a_{N-1},\ua}|0\ket=
\tilde{\ve}_{\a_{1}..\a_{N-1}}
\sum_{i_{k}>..>i_{1}=N,\neq \a_{1}}^{2N-2}
\hat{S}^{-}_{i_{k}}..\hat{S}^{-}_{i_{1}}
c^{\dag}_{N,\ua}..c^{\dag}_{2N-2,\ua}|0\ket+...
\end{eqnarray}
Since the $c^{\dag}_{\a_{1},\ua}$ creation operator commutes with all 
the lowering spin operators coming from the sum over 
$i_{k}>..>i_{1}$,  
one of the indices $n_{N-k-1}>..>n_{1}$ is constrained to be 
$\a_{1}$ and the third row of Eq.(\ref{fsp}) can be rewritten as 
\begin{eqnarray}
\tilde{\ve}_{\a_{1}..\a_{N-1}}
\sum_{n_{N-k-1}>..>n_{2}=N,\neq \a_{1}}^{2N-2}
\sum_{i_{k}>..>i_{1}=N,\neq \a_{1}}^{2N-2}
\bra 0|c_{2N-2,\da}..c_{N,\da}
\hat{S}^{-}_{n_{N-k-1}}..\hat{S}^{-}_{n_{1}}
\hat{S}^{-}_{i_{k}}..\hat{S}^{-}_{i_{1}}
c^{\dag}_{N,\ua}..c^{\dag}_{2N-2,\ua}|0\ket
=\nonumber \\ =
\left(\begin{array}{c} N-2 \\ k \end{array}\right)
\tilde{\ve}_{\a_{1}..\a_{N-1}}.
\end{eqnarray}
Substituting these results in Eq.(\ref{fsp}) one obtains 
\begin{equation}
\fbox{$
Y({\mathbf r})=\frac{F_{N}}{|{\cal S}_{hf}|^{2}}=
\frac{1}{4}-\frac{1}{|{\cal S}_{hf}|^{2}}[D+E\times (4N-4)]$}
\label{oddips}
\end{equation}
where we have taken into account that ${\mathrm Det}[T({\mathbf r})]=-1$ for odd 
$x+y$ and the constant $F_{N}$ is given by
\begin{equation}
\fbox{$
F_{N}=\frac{|{\cal S}_{hf}|^{2}}{{\cal N}}
\sum_{k=1}^{N-2}f_{k}f_{N-k-1}
\left(\begin{array}{c} N-2 \\ k-1 \end{array}\right)
\left(\begin{array}{c} N-2 \\ k \end{array}\right)$}\;\;.
\end{equation}
In the last equality of Eq.(\ref{oddips}) we have used 
$F_{N}=|{\cal S}_{hf}|^{2}/4-D-E\times (4N-4)$ which is a direct 
consequence of the sum rule for the charge correlation function, see 
Eq.(\ref{srgc}).

}

\end{document}